\documentclass{class/tlp}

\usepackage[utf8]{inputenc}
\usepackage[T1]{fontenc}
\usepackage{amsthm, amsmath, amsfonts, amssymb, mathtools, mathabx, stmaryrd}
\usepackage[bb=boondox]{mathalfa}
\usepackage{graphicx}
\usepackage{multicol}
\usepackage{ifdraft}
\usepackage{ebproof}
\usepackage{algorithm}
\usepackage{algorithmicx}
\usepackage{xspace}
\usepackage[noend]{algpseudocode}
\usepackage{tikz, pgf}
\usepackage{hyperref}
\usepackage{url}
\usepackage[figbotcap]{subfigure}
\usepackage{lib/latexutils/abbrev}
\usepackage{lib/latexutils/sidenote}
\usepackage{lib/latexutils/symbols}
\usepackage{lib/latexutils/theorems} 
\usepackage{adjustbox}

\usepackage[frozencache,cachedir=_minted-preprint_finalizecache]{minted}

\newcommand{\prolog}[1]{\mintinline{prolog}{#1}}
\newcommand{\haskell}[1]{\mintinline{haskell}{#1}}

\newcommand{\python}[1]{\mintinline{python}{#1}}

\setminted{
  linenos=true,
  autogobble
}

\ebproofset{
  right label template=\textsc{\inserttext}
}
\ebproofnewstyle{small}{
  separation = 1em, rule margin = .5ex,
  template = \footnotesize$\inserttext$,
  right label template=\footnotesize\textsc{\inserttext}
}

\usetikzlibrary{cd, arrows.meta, decorations.pathmorphing}
\tikzset{
  >=Kite,
  invisible/.style={opacity=0},
  visible on/.style={alt={#1{}{invisible}}},
  alt/.code args={<#1>#2#3}{%
      \alt<#1>{\pgfkeysalso{#2}}{\pgfkeysalso{#3}}%
  }
}
\tikzcdset{
  arrow style=tikz,
  diagrams={>=Kite},
  ampersand replacement=\&,
  labels=description
}

\algrenewcommand\algorithmicindent{1.0em}%
\algrenewcommand\algorithmicfunction{\textbf{fn}}
\algnewcommand\algorithmicyield{\textbf{yield}\ }
\algnewcommand\algorithmiccontinue{\textbf{continue}}
\algnewcommand\algorithmicbreak{\textbf{break}}

\begin{document}
  \lefttitle{S. Rechenberger, T. Frühwirth}

\jnlPage{1}{28}
\jnlDoiYr{2025}
\doival{10.1017/xxxxx}

\title[FreeCHR]{FreeCHR – An Algebraic Framework for CHR Embeddings}

\begin{authgrp}
\author{\gn{Sascha} \sn{Rechenberger} and \gn{Thom} \sn{Frühwirth}}
\affiliation{%
    Institute of Software Engineering and Programming Languages\\
    Ulm University, 89069 Ulm, Germany\\
    \email{\{sascha.rechenberger,thom.fruehwirth\}@uni-ulm.de}
}
\end{authgrp}


\maketitle
  \begin{abstract}
  We introduce the framework \emph{FreeCHR} which formalizes the embedding of \emph{Constraint Handling Rules} (CHR) into a host language, using the concept of \emph{initial algebra semantics} from category theory\footnote{This is an extended version of a paper presented at the 2023 RuleML+RR Conference.
  The authors are grateful to Anna Fensel, Ana Ozaki and Ahmet Soylu, the conference program co-chairs, for encouraging us to submit the paper to the journal Theory and Practice of Logic Programming and for their help in overseeing the review process.}.
  We hereby establish a high-level implementation scheme for CHR as well as a common formalization for both theory and practice.
  We propose a lifting of the syntax of CHR via an endofunctor in the category \textbf{Set} and a lifting of the very abstract operational semantics of CHR into FreeCHR, using the free algebra, generated by the endofunctor.
  We give proofs for soundness and completeness \WRT its original definition.
  We also propose a first abstract execution algorithm and prove correctness \WRT the operational semantics.
  Finally, we show the practicability of our approach by giving two possible implementations of this algorithm in Haskell and Python. 
  Under consideration in Theory and Practice of Logic Programming (TPLP).
\end{abstract}


  \ifdraft{\tableofcontents}

  \section{Introduction}\label{sec::freechr:introduction}
\emph{Constraint Handling Rules} (CHR) is a rule-based programming language that is designed to be embedded into a general-purpose language.
Having a CHR implementation available enables software developers to solve problems in a declarative and elegant manner.
Aside from the obvious task of implementing constraint solvers (\cite{fruehwirth2006complete,dekoninck2006inclp}), it has been used, \EG to solve scheduling problems (\cite{abdennadher2000university}) implement concurrent and multi-agent systems (\cite{thielscher2002reasoning,thielscher2005flux,lam2006agent,lam2007concurrent}).
In general, CHR is ideally suited for any problem that involves the transformation of collections of data, as programs consist of a set of rewriting rules, hiding away the process of finding suitable candidates for rule application.
Hereby, we get a purely declarative representation of the algorithm without the otherwise necessary boilerplate code.

The literature on CHR as a formalism consists of a rich body of theoretical work concerning CHR, including a rigorous formalization of its declarative and operational semantics (\cite{fruehwirth2009constraint,sneyers2010time,fruehwirth2015constraint}), relations to other rule-based formalisms (\EG \cite{betz2007relating} and \cite{gall2018operational}) and results on properties like confluence (\cite{christiansen2015confluence}).

Implementations of CHR exist for a number of languages, such as Prolog (\cite{schrijvers2004leuven}), C (\cite{wuille2007cchr}), C++ (\cite{barichard2024chr}), Haskell (\cite{chin2008typesafe,lam2007concurrent}), JavaScript (\cite{nogatz2018chr}) and Java (\cite{abdennadher2002jack,vanweert2005leuven,ivanovic2013implementing,wibiral2022javachr}).

While the implementations adhere to the formally defined operational semantics, they are not direct implementations of a common formal model.
Therefore, the two aspects of CHR (formalism and programming language) are not strictly connected with each other and
there is hence no strict guarantee that the results on the formalism CHR are applicable on the programming language CHR.
Although, such a strict connection is probably not entirely achievable (unless we define and use everything inside a proof assistant like \emph{Coq} or \emph{Agda}), it is desirable to have both formal definition and implementation as closely linked as possible.
Additionally, to being able to directly benefit from theoretical results, implementors of CHR embeddings and users of the CHR language can also use the formally defined properties to define more profound tests for their software.

Another apparent issue within the CHR ecosystem is that many of the implementations of CHR are currently unmaintained.
Although some of them are mere toy implementations, others might have been of practical use.
One example is JCHR (\cite{vanweert2005leuven}) which would be a useful tool if it was kept on par with the development of Java, especially with modern build tools like \emph{Gradle}.
Having a unified formal model from which every implementation is derived could ease contributing to implementations of CHR as it provides a strict documentation and description of the system a priori.
Also, different projects might even be merged which would prevent confusion due to multiple competing, yet very similar implementations, as it can be observed in the \emph{miniKanren} ecosystem (\EG there exist about 20 implementations of \emph{miniKanren} dialects only for Haskell\footnote{\url{https://minikanren.org/\#implementations}}).

A third major issue is that many implementations, like the aforementioned JCHR or CCHR (\cite{wuille2007cchr}), are implemented as external EDSL, \IE they rely on a separate compiler which translates CHR code into code of the host language.
This makes it significantly less convenient to use and hence also less likely to be adopted into the standard distribution of the host language.
This is somewhat demonstrated by the \emph{K.U.Leuven CHR system} which is implemented as a library in Prolog and distributed as a standard package with \emph{SWI-Prolog}\footnote{\url{https://www.swi-prolog.org/pldoc/man?section=chr}}, or by the library \emph{core.logic}\footnote{see \url{https://github.com/clojure/core.logic}} which implements \emph{miniKanren} for the LISP dialect \emph{Clojure}.

To solve the three issues stated above, we introduce the framework \emph{FreeCHR} which formalizes the embedding of CHR, using \emph{initial algebra semantics}.
This common concept in functional programming is used to inductively define languages and their semantics (\cite{hudak1998modular,johann2007initial}).
FreeCHR provides both a guideline and high-level architecture to implement and maintain CHR implementations across host languages and a strong connection between the practical and formal aspects of CHR.
To our knowledge, the presented execution algorithm and its implementations, although simple, will be the first implementations of CHR for which there are formal proofs of correctness.
Also, by FreeCHR-instances being internal embeddings, we get basic tooling like syntax highlighting and type-checking for free (\cite{fowler2011domainspecific}) and are able to implement CHR as a library, without the need for any external tooling like source-to-source compilers.

Ultimately, the framework shall serve a fourfold purpose by providing \begin{itemize}
    \item a general guideline on how to implement a CHR system in modern high-level languages,
    \item a guideline for future maintenance of FreeCHR instances,
    \item a common framework for both formal considerations and practical implementations
    \item and a framework for the definition and verification of general criteria of correctness.
\end{itemize}

In this work, we will give first formal definitions of FreeCHR, upon which we will build our future work.
\autoref{sec::freechr:preliminaries} will provide the necessary background and intuitions.
\autoref{sec::freechr:non-herbrand-chr} introduces the syntax and semantics of Constraint Handling Rules and generalizes them to non-Herbrand domains.
\autoref{sec::freechr:freechr} introduces the framework FreeCHR.
\autoref{sec::freechr:fr:syntax} lifts the syntax of CHR programs to a \textbf{Set}-endofunctor and introduces the free algebra, generated by that functor,
\autoref{sec::freechr:fr:operational-semantics} lifts the \emph{very abstract} operational semantics $\omega_a$ of CHR to the  \emph{very abstract} operational semantics $\omstara$ of FreeCHR
and \autoref{sec::freechr:fr:correctness} proves soundness and completeness of $\omstara$ \WRT $\oma$.

\autoref{sec::freechr:instance:very-abstract} introduces a simple execution algorithm for a formal instance with \emph{very abstact} operational semantics and proves their correctness. This is the second main contribution of this paper, as it gives an example use of a proven correct guideline to implement FreeCHR.

This is finally demonstrated with the two case studies in \autoref{sec::freechr:instance:very-abstract:case-studies}. The section provides two implementations of the defined instance in Haskell and Python. The languages were chosen, as they can be considered to be on the opposite sides of a static-dynamic spectrum of programming languages.
They will hence give a good intuition of how the host language flavors FreeCHR.

Finally, \autoref{sec::freechr:limitations} discusses the limitations of FreeCHR as presented in this paper,
\autoref{sec::freechr:related-work} discusses related work,
\autoref{sec::freechr:future-work} will give an overview over planned and ongoing future work and \autoref{sec::freechr:conclusion} concludes the paper.

  \section{Preliminaries}\label{sec::freechr:preliminaries}

\subsection{Endofunctors and F-algebras}
In this section, we want to introduce endofunctors and $F$-algebras.
Both concepts are taken from category theory and will be introduced as instances in the category of sets \textbf{Set}.

We do not assume any previous knowledge of category theory, but to readers more interested in the topic in general we recommend \cite{milewski2019category} as introductory literature.

\subsubsection{Basic definitions}
The \emph{disjoint union} of two sets $A$ and $B$
\begin{align*}
  A \sqcup B = \left\{l_A(a) \mid a \in A\right\} \cup \left\{l_B(b) \mid b \in B\right\}
\end{align*}
is the union of both sets with additional labels $l_A$ and $l_B$ added to the elements to keep track of the origin set of each element.
We will also use the labels $l_A$ and $l_B$ as \emph{injection} functions $l_A : A \rightarrow A \sqcup B$ and $l_B : B \rightarrow A \sqcup B$ which construct elements of $A \sqcup B$ from elements of $A$ or $B$, respectively.

For two functions $f : A \rightarrow C$ and $g : B \rightarrow C$, the function
\begin{align*}
  \left[f,g\right] &: A \sqcup B \rightarrow C\\
  \left[f,g\right](l(x)) &=\left\{\begin{matrix*}[l]
    f(x), & & \mbox{if}\ l = l_A \\
    g(x), & & \mbox{if}\ l = l_B \\
  \end{matrix*}\right.
\end{align*}
is called a \emph{case analysis} function of the disjoint union $A \sqcup B$.
It can be understood as a formal analog to a \texttt{case-of} expression.
Furthermore, we define two functions
\begin{multicols}{2}
  \noindent
  \begin{align*}
    f \sqcup g &: A \sqcup B \rightarrow A' \sqcup B'\\
    (f \sqcup g)(l(x)) &= \left\{
      \begin{matrix*}[l]
        l_{A'}(f(x)), & &\mbox{if}\ l = l_A \\
        l_{B'}(g(x)), & &\mbox{if}\ l = l_B
      \end{matrix*}
    \right.
  \end{align*}
  \begin{align*}
    f \times g &: A \times B \rightarrow A' \times B'\\
    (f \times g)(x,y) &= (f(x), g(y))
  \end{align*}
\end{multicols}
\noindent
which lift two functions $f : A \rightarrow A'$ and $g : B \rightarrow B'$ to the disjoint union and the Cartesian product, respectively.

\subsubsection{Endofunctors}
A \textbf{Set}-endofunctor\footnote{Since we only deal with endofunctors in \textbf{Set} we will simply call them \emph{functors}.} $F$ maps all sets $A$ to sets $F A$ and all functions $f : A \rightarrow B$ to functions $F f : F A \rightarrow F B$, 
such that $F \id_A = \id_{F A}$ and $F (g \circ f) = F g \circ F f$, where $\id_X(x) = x$ is the identity function on a set $X$\footnote{We will omit the index of $\id$ if it is clear from the context.}.
A signature $\Sigma = \left\{\sigma_1/a_1,...,\sigma_n/a_n\right\}$, where $\sigma_i$ are operators and $a_i$ their arity, generates a functor
\begin{align*}
  F_{\Sigma} X = \bigsqcup_{\sigma/a\in\Sigma} X^{a} && 
  F_{\Sigma} f = \bigsqcup_{\sigma/a\in\Sigma} f^{a}
\end{align*}
with $X^0 = \mathbb{1}$ and $f^0 = \id_{\mathbb{1}}$, where $\mathbb{1}$ is a singleton set.
Such a functor $F_{\Sigma}$ models \emph{flat} (\IE not nested) terms over the signature $\Sigma$.

\begin{example}
  The signature
  \begin{align*}
    \Gamma = \left\{\mathbf{0}/0, \oplus/2, \mathbf{1}/0, \otimes/2\right\}
  \end{align*}
  models two constants $\mathbf{0}$ and $\mathbf{1}$, as well as two binary operators $\oplus$ and $\otimes$.
  It generates the functor
  \begin{align*}
    F_{\Gamma} X &= \mathbb{1} \sqcup X \times X \sqcup \mathbb{1}  \sqcup X \times X\\
    F_{\Gamma} f &= \id_{\mathbb{1}} \sqcup f \times f \sqcup \id_{\mathbb{1}} \sqcup f \times f
  \end{align*}
\end{example}

\subsubsection{$F$-algebras}
Since an endofunctor $F$ defines the syntax of terms, an evaluation function $\alpha : F A \rightarrow A$ defines the \emph{semantics} of terms.
We call such a function $\alpha$, together with its \emph{carrier} $A$, an $F$-algebra $(A, \alpha)$.

If there are two $F$-algebras $\left(A, \alpha\right)$ and $\left(B, \beta\right)$ and a function $h : A \rightarrow B$, we call $h$ an \emph{$F$-algebra homomorphism} \IFF $h \circ \alpha = \beta \circ F h$, i.e., $h$ preserves the structure of $\left(A, \alpha\right)$ in $\left(B, \beta\right)$ when mapping $A$ to $B$.
In this case, we also write $h : \left(A,\alpha\right) \rightarrow \left(B, \beta\right)$.

A special $F$-algebra is the \emph{free $F$-algebra} $F^{\star} = (\mu F, \cons_F)$, for which there is a homomorphism $\kata{\alpha} : F^{\star} \rightarrow \left(A, \alpha\right)$ for any other algebra $\left(A, \alpha\right)$.
We call those homomorphisms $\kata{\alpha}$ \emph{$F$-catamorphisms}.
The functions $\kata{\alpha}$ encapsulate structured recursion on values in $\mu F$ with the semantics defined by the function $\alpha$ which is itself only defined on flat terms.
The carrier of $F^{\star}$, with $\mu F = F \mu F$, is the set of inductively defined values in the shape defined by $F$.
The function $\cons_F : F \mu F \rightarrow \mu F$ inductively constructs the values in $\mu F$.

\begin{example}
  The initial $F_{\Gamma}$-algebra
  \begin{align*}
    \left(\mu F_{\Gamma}, \left[\mathbf{0}, \oplus, \mathbf{1}, \otimes\right]\right)
  \end{align*}
  can be used to construct the nested expressions like $\mathbf{1} \oplus (\mathbf{1} \otimes \mathbf{0})$ of $\mu F_{\Gamma}$.
  The $F_\Gamma$ algebras
  \begin{align*}
    \left(\tbool, \left[\bfalse, \bor, \btrue, \band\right]\right) && \left(\tnatzero, \left[0, +, 1, \cdot\right]\right)
  \end{align*}
  give to $F_{\Gamma}$ the semantics of Boolean and arithmetic expressions, respectively.
  Using the catamorphism $\kata{\left[\bfalse, \bor, \btrue, \band\right]} : \mu F_{\Gamma} \rightarrow \tbool$,
  we can evaluate the expression from above as
  \begin{align*}
    &\kata{\left[\bfalse, \bor, \btrue, \band\right]}(\mathbf{1} \oplus (\mathbf{1} \otimes \mathbf{0}))\\
    = &\kata{\left[\bfalse, \bor, \btrue, \band\right]}(\mathbf{1}) \bor \kata{\left[\bfalse, \bor, \btrue, \band\right]}(\mathbf{1} \otimes \mathbf{0})\\
    = &\btrue \bor \kata{\left[\bfalse, \bor, \btrue, \band\right]}(\mathbf{1}) \band \kata{\left[\bfalse, \bor, \btrue, \band\right]}(\mathbf{0})\\
    = &\btrue \bor (\btrue \band \bfalse)\\
    = &\btrue \bor \bfalse\\
    = &\btrue
  \end{align*}
  Analogously, if we use $\kata{\left[0, +, 1, \cdot\right]} : \mu F_{\Gamma} \rightarrow \tnatzero$, we get
  \begin{align*}
    &\kata{\left[0, +, 1, \cdot\right]}(\mathbf{1} \oplus (\mathbf{1} \otimes \mathbf{0}))\\
    = &\kata{\left[0, +, 1, \cdot\right]}(\mathbf{1}) + \kata{\left[0, +, 1, \cdot\right]}(\mathbf{1} \otimes \mathbf{0})\\
    = &1 + \kata{\left[0, +, 1, \cdot\right]}(\mathbf{1}) \cdot \kata{\left[0, +, 1, \cdot\right]}(\mathbf{0})\\
    = &1 + (1 \cdot 0)\\
    = &1 + 0\\
    = &1
  \end{align*}

  Technically, the constants $\mathbf{0}$, $\mathbf{1}$, $\bfalse$, $\btrue$, $0$ and $1$ are supposed to be functions from the singleton set $\mathbb{1}$ to the respective carrier.
  Since $\mathbb{1}$ only has one element, there is essentially no difference between such a function and the constant itself.
  We will hence use the constant directly.
\end{example}

\subsection{Labelled transition systems}\label{sec::lts}

In this section, we will lay out our definition and notation of \emph{labelled transition systems} and notions of soundness and completeness.

\begin{definition}[Labelled transition system]
  A \emph{labelled transition system} (LTS) is a structure $\langle S, A, R \rangle$,
  where $S$ is the set of states called the \emph{domain} of the system, $A$ is a set of \emph{labels} and
  $R \in S \times A \times S$ a ternary \emph{transition relation}.
\end{definition}

For an LTS $\omega = \langle S, A, \mapsto \rangle$, we write $s \xmapsto{a} s'$, if $(s, a, s') \in \left(\mapsto\right)$ and $s_0 \xmapsto{a_1} s_1 \xmapsto{a_2} s_2 \xmapsto{a_3} ... \xmapsto{a_n} s_n$ for $s_0 \xmapsto{a_1} s_1 \wedge s_1 \xmapsto{a_2} s_2 \wedge ... \wedge s_{n-1} \xmapsto{a_n} s_n$.
Also, we will write $s_1 \xmapsto{a}^{*} s_n$ if there is an $a \in A$, such that $s_1 \xmapsto{a} s_2 \wedge ... \wedge s_{n-1} \xmapsto{a} s_n$, for $n \geq 0$.

\begin{definition}[$\theta$-Soundness]
  Given two labelled transition systems $\omega = \langle S, A, \mapsto \rangle$ and $\omega' = \langle S, A', \hookrightarrow \rangle$.
  For function $\theta : A \rightarrow A'$, we call $\omega$ \emph{$\theta$-sound} \WRT $\omega'$ \IFF
  \begin{align*}
    s \xmapsto{a} s' \Longrightarrow s \xhookrightarrow{\theta(a)} s' 
  \end{align*}
  for all $s, s' \in S$ and $a \in A$ \footnote{We will generally just use the symbol $\mapsto$ for the relation if the respective transition system is clear from the context.}
\end{definition}

$\theta$-soundness means that any transition defined by $\omega$ is also defined by $\omega'$, when mapping labels with $\theta$.
This is typically used to verify a more constrained system ($\omega$) against a more general one ($\omega'$).

Dually, there exists a concept of \emph{completeness}.
\begin{definition}[$\theta$-Completeness]
  Given two labelled transition systems $\omega = \langle S, A, \mapsto \rangle$ and $\omega' = \langle S, A', \hookrightarrow \rangle$.
  For a function $\theta : A \rightarrow A'$, we call $\omega$ \emph{$\theta$-complete} \WRT $\omega'$ \IFF
  \begin{align*}
    s \xmapsto{a} s' \Longleftarrow s \xhookrightarrow{\theta(a)} s' 
  \end{align*}
  for all $s, s' \in S$ and $a \in A$.
\end{definition}

$\theta$-completeness means that $\omega$ defines all transitions that are defined by $\omega'$, but using only the mapped labels of $\omega$.

We will also need a slightly relaxed notion of soundness.

\begin{definition}[$\theta$-Soudness up to repeated transition]
  Given two labelled transition systems $\omega = \langle S, A, \mapsto \rangle$ and $\omega' = \langle S, A', \hookrightarrow \rangle$.
  For a function $\theta : A \rightarrow A'$, we call $\omega$ \emph{$\theta$-sound} \emph{up to repeated transition} \WRT $\omega'$ \IFF
  \begin{align*}
    s \xmapsto{a} s' \Longrightarrow s \xhookrightarrow{\theta(a)}^{*} s'
  \end{align*}
  for all $s, s' \in S$, $a \in A$ and $l \geq 0$.
\end{definition}

This definition says that an LTS is also $\theta$-sound \WRT another if we can accomplish a transition in one LTS with zero or more transitions in the other.

  \section{CHR over non-Herbrand domains}\label{sec::freechr:non-herbrand-chr}
The first implementations of CHR were embedded into the logical programming language Prolog,
where terms like \prolog{3+4} or \prolog{f(a,b,c)} are not evaluated, as is the case in most other programming languages,
but interpreted as themselves.
This is called the \emph{Herbrand} interpretation of terms.
Since we want to embed CHR in any programming language, we need to generalize the language to non-Herbrand interpretations of terms.
We will formalize this, using initial algebra semantics.

\subsection{Host language}
We first define a \emph{data type} in the host language.
A data type determines the syntax and semantics of terms via a functor $\Lambda_T$ and an algebra $\tau_T$.
The fixed point $\mu\Lambda_T$ contains terms that are inductively defined via $\Lambda_T$ and the catamorphism $\kata{\tau_T}$ evaluates those terms to values of $T$.

\begin{definition}[Data types]\label{def::lan:data-type}
  A \emph{data type} is a triple $\langle T, \Lambda_{T}, \tau_T\rangle$, where $T$ is a set, $\Lambda_T$ a functor and $\left(T, \tau_T\right)$ a $\Lambda_T$-algebra.
\end{definition}
We write $t \equiv_{T} t'$ for $t \in \mu\Lambda_T$ and $t' \in T$ \IFF $\kata{\tau_T}(t) = t'$. 

\begin{example}[Boolean data type]
  The signature
  \begin{align*}
    \Sigma_{\tbool} =& \left\{(n \leq m)/0 \mid n,m\in\mathbb{N}_0\right\}
      \cup \left\{(n < m)/0 \mid n,m\in\mathbb{N}_0\right\}
      \cup \left\{\band/2, \btrue/0, \bfalse/0\right\}
  \end{align*}
  defines Boolean terms\footnote{We will generally overload symbols like $\bfalse$, $\btrue$, $\band$, $\bnot$, ..., if their meaning is clear from the context.}.
  $\Sigma_{\tbool}$ generates the functor
  \begin{align*}
    \Lambda_{\tbool} X =&\ \mathbb{N}_0 \times \mathbb{N}_0
     \sqcup\ \mathbb{N}_0 \times \mathbb{N}_0
     \sqcup\ X \times X \sqcup \mathbb{1} \sqcup \mathbb{1}
  \end{align*}
  the fixed point, $\mu\Lambda_{\tbool}$, of which is the set of valid nested Boolean terms like $\left(0 < 4 \band 4 \leq 6\right)$.
  Let $\langle \tbool, \Lambda_{\tbool}, \tau_{\tbool} \rangle$, with $\tbool = \left\{\btrue, \bfalse\right\}$, be a data type. 
  If we assume $\tau_{\tbool}$ to implement the usual semantics for Boolean terms and comparisons, $\left(0 < 4 \band 4 \leq 6\right)$ will evaluate as 
  \begin{align*}
    \kata{\tau_{\tbool}}(0 < 4 \band 4 \leq 6) = \kata{\tau_{\tbool}}(0 < 4) \band \kata{\tau_{\tbool}}(4 \leq 6) = \btrue \band \btrue = \btrue
  \end{align*}
\end{example}

For a set $T$, both $\Lambda_T$ and $\tau_T$ are determined by the host language which is captured by the next definition.

\begin{definition}[Host environment]\label{def::lan:host-environment}
  A mapping
  \begin{align*}
    \mathcal{L} T = \langle T, \Lambda_T, \tau_T \rangle
  \end{align*}
  where $\langle T, \Lambda_T, \tau_T \rangle$ is a data type, is called a \emph{host environment}.
\end{definition}

A host environment is implied by the host language (and by the program, the CHR program is part of) and assigns to a set $T$ a data type, effectively determining syntax and semantics of terms that evaluate to values of $T$.

\subsection{Embedding CHR}
With the formalization of our host environment, we can define the syntax and semantics of CHR.

\begin{definition}[CHR programs]\label{def::chr:syn:program}
  CHR programs are sets of multiset-rewriting rules of the form 
  \begin{align*}
    &N\ @\ K\ \setminus\ R\ \Longleftrightarrow\ G\ |\ B
  \end{align*}
  For a set $C$, called the \emph{domain} of the program, for which there is a data type $\mathcal{L} C = \langle C, \Lambda_C, \tau_C \rangle$,
  $K, R \in \flist C$ are called the \emph{kept} and \emph{removed head}, respectively.
  The backslash ($\setminus$) is used to separate the kept and removed part of the head.
  Either $K$ or $R$ must be non-empty.
  If either head part is empty, the separator is omitted as well.
 
  $\flist X = \bigcup_{i\in\tnatzero} X^i$ maps a set $X$ to the set of lists over X, with $X^0 = \eps$ being the empty sequence,
  and functions $f : A \rightarrow B$ to functions\begin{align*}
    \flist f \left(a_1, ..., a_n\right) = \left(f(a_1), ..., f(a_n)\right)
  \end{align*}
  
  The optional $G \in \mu\Lambda_{\tbool}$ is called the \emph{guard}.
  If $G$ is omitted, we assume $G \equiv_{\tbool} \btrue$ and omit the pipe ($|$) which separates the guard from the body as well.
  $B \in \fmultiset\mu\Lambda_C$ is called the \emph{body}.
  The functor $\fmultiset$ maps sets $X$ to the set of multisets over $X$ and functions $f : X \rightarrow Y$ to functions
  \begin{align*}
    \fmultiset f \left(\left\{a_1, a_2, ...\right\}\right) = \left\{f(a_1), f(a_2), ...\right\} 
  \end{align*}
  $N$ is an optional name for the rule, which is generally used for debugging and tracing.
  The $@$ symbol is used to separate the name of the rule from the rule itself.
  It is also omitted if the rule is not given a name.
\end{definition}

The members of the kept and removed head are matched against values of the domain $C$.
The guard $G$ is a term that can be evaluated to a Boolean value.
The body $B$ is a multiset over terms which can be evaluated to values of $C$.
This includes any call of functions or operators which evaluate to Boolean, or values of $C$, respectively.

In literature, rules with an empty kept head are called \emph{simplification} rules and rules with an empty removed head \emph{propagation} rules.
Rules which have both a removed and kept head are called \emph{simpagation} rules.

\autoref{def::chr:syn:program} corresponds to the \emph{positive range-restricted ground} segment of CHR which is commonly used as the target for embeddings of other (rule-based) formalisms (\EG colored Petri nets as by \cite{betz2007relating}) into CHR \cite[Chapter 6.2]{fruehwirth2009constraint}.
\emph{Positive} means that the body of the rule contains only user constraints (\IE values from $C$) which guarantees that computations do not fail.
\emph{Range-restricted} means that instantiating all variables of the head ($K$ and $R$) will ground the whole rule.
This also maintains the \emph{groundness} of the segment of CHR which requires that the input and output of a program are ground.
$\prgc$ denotes the set of all such programs over a domain $C$.

\begin{example}[Euclidean algorithm]\label{ex::freechr:pre:chr:gcd}
  The program $\textsc{gcd}=\left\{\mathit{zero}, \mathit{subtract}\right\}$\footnote{We will typically use the rule name as a symbol for the rule as a whole.}
  \begin{align*}
    \mathit{zero}\ &@\ 0\ \Leftrightarrow\ \emptyset\\
    \mathit{subtract}\ &@\ N\ \setminus\ M\ \Leftrightarrow\ 0 < N \band 0 < M \band N \leq M\ |\ M-N
  \end{align*}
  computes the greatest common divisor of a collection of natural numbers.
  The first rule removes all zeros from the collection.
  For any pair of numbers $N$ and $M$ greater $0$ and $N \leq M$, the second rule replaces $M$ by $M-N$.
  Note that we omitted the kept head and guard of the $zero$ rule.
\end{example}

\begin{definition}[$C$-instances of rules]\label{def::chr:syn:pro:instance}
  For a \emph{positive range-restricted} rule
  \begin{align*}
    R\ @\ k_1,...,k_n\ \setminus\ r_{1},...,r_{m}\ \Leftrightarrow\ G\ |\ B
  \end{align*}
  with universally quantified variables $v_1,...,v_l$,
  and a data type $\mathcal{L} C = \langle C, \Lambda_{C}, \tau_{C} \rangle$, 
  we call the set
  \begin{align*}
    \Gamma_{C}(R) =
      \{\ &\left(R\ @\ k_1\sigma,...,k_n\sigma\ \setminus\ r_{1}\sigma,...,r_{m}\sigma\ \Leftrightarrow\ G\sigma\ |\ \fmultiset\kata{\tau_{C}}(B\sigma) \right) \\
      \mid\ &\sigma\ \mbox{instantiates all variables $v_1$, ..., $v_l$},\\
      \ &k_1\sigma,...,k_n\sigma, r_1\sigma,...,r_m\sigma \in C,\\
      \ &G \sigma \in \mu\Lambda_{\tbool},\\
      \ &B \sigma \in \fmultiset\mu\Lambda_C\ \}
  \end{align*}
  the \emph{$C$-grounding} of $R$.
  Analogously, for a set $\mathcal{R}$ of rules, $\Gamma_C(\mathcal{R}) = \bigcup_{R \in \mathcal{R}} \Gamma_C(R)$ is the $C$-grounding of $\mathcal{R}$.
  An element $r' \in \Gamma_{C}(R)$ (or $\Gamma_C(\mathcal{R})$ respectively) is called a \emph{$C$-instance} of a rule $R \in \mathcal{R}$.
\end{definition}

A C-instance (or grounding) is obtained, by instantiating all variables and evaluating the then ground terms in the body of the rule, using the $\Lambda_C$-catamorphism $\kata{\tau_C}$.
The functor $\fmultiset$ is used to lift $\kata{\tau_C}$ into the multiset.

\begin{example}\label{ex::freechr:nhc:body-instance}
Given a body $\left\{M - N\right\}$ and a substitution $\sigma = \left\{N \mapsto 4, M \mapsto 6\right\}$, the body is instantiated like
\begin{align*}
  \fmultiset\kata{\tau_C}(\left\{M-N\right\}\sigma) = \fmultiset\kata{\tau_C}(\left\{6-4\right\}) = \left\{\kata{\tau_C}(6-4)\right\} = \left\{2\right\}
\end{align*}
\end{example}
With \autoref{ex::freechr:nhc:body-instance}, we can also easily see that if we use a data type $\mathcal{L}\mu\Lambda_C = \langle \mu\Lambda_C, \Lambda_C, \cons_{\Lambda_C} \rangle$ we get the Herbrand interpretation of terms over $C$.
Hence, \FI an expression $\left(3+4\right) \in \mu\Lambda_{\mathbb{N}_0}$ is evaluated to itself, as it is the case in Prolog.

\begin{example}[$C$-instances]\label{ex::chr:syn:pro:instance}
  If we instantiate the rule
  \begin{align*}
    subtract\ &@\ N\ \setminus\ M\ \Leftrightarrow\ 0 < N \band 0 < M \band N \leq M\ |\ M-N
  \end{align*} 
  with $\sigma_1 = \left\{N \mapsto 4, M \mapsto 6\right\}$ and $\sigma_2 = \left\{N \mapsto 0,M \mapsto 6\right\}$, respectively,
  we get the $\mathbb{N}_0$-instances
  \begin{align*}
    (subtract) \sigma_1 = subtract\ &@\ 4\ \setminus\ 6\ \Leftrightarrow\ 0 < 4 \band 0 < 6 \band 4 \leq 6\ |\ 2\\
    (subtract) \sigma_2 = subtract\ &@\ 0\ \setminus\ 6\ \Leftrightarrow\ 0 < 0 \band 0 < 6 \band 0 \leq 6\ |\ 6
  \end{align*}
  Both instances are elements of the $\mathbb{N}_0$-grounding $\Gamma_{\mathbb{N}_0}\left(\textsc{gcd}\right)$ of the program in \autoref{ex::freechr:pre:chr:gcd}.
\end{example}

Classically, the guard $G$ contains constraints which are defined \WRT a constraint theory $\mathcal{CT}$.
We typically write $\mathcal{CT} \models G$ \footnote{As we only work with ground values, we do not need to quantify any variables.} to denote that the guard is satisfiable \WRT $\mathcal{CT}$ and $\mathcal{CT} \models \bnot G$ otherwise.
Since in our case $G \in \mu\Lambda_{\tbool}$, $\mathcal{CT}$ is essentially $\tau_{\tbool}$, as it determines the semantics of Boolean terms. We thus write
\begin{align*}
  \tau_{\tbool} \models G \Longleftrightarrow G \equiv_{\tbool} \btrue &&
  \tau_{\tbool} \models \bnot G \Longleftrightarrow G \equiv_{\tbool} \bfalse
\end{align*}
Note that we always need a data type $\mathcal{L} \tbool$. In Prolog, \FI $\tbool$ corresponds to the set $\left\{\mbox{\prolog{true}}, \mbox{\prolog{false}}\right\}$, representing successful or failed computations, respectively.

Finally, the operational semantics of CHR is defined as a state transition system where the states are multisets\footnote{There may be some additional decoration in more refined operational semantics.} over the elements of $C$.

\begin{definition}[\emph{Very abstract} operational semantics of CHR]\label{def::chr:sem:op:very-abstract}
  The \emph{very abstract} operational semantics of CHR programs over a domain $C$ is given by the labelled transition system
  \begin{align*}
    \omega_a = \langle \fmultiset C, \prgc, \mapsto \rangle
  \end{align*}
  where the transition relation $\left(\mapsto\right)$ is defined by the inference rules in \autoref{fig-210-omega-a-semantics}.
  \begin{figure}[t]
    \begin{align*}
      \begin{prooftree}
        \infer0[empty]{S \xmapsto{R} S}
      \end{prooftree}
      &&
      \begin{prooftree}
        \hypo{r \in R}
        \hypo{S \xmapsto{\left\{r\right\}} S'}
        \hypo{S' \xmapsto{R} S''}
        \infer3[step]{S \xmapsto{R} S''}
      \end{prooftree}
    \end{align*}

    \begin{align*}
      \begin{prooftree}
        \hypo{R@c_{1},...,c_{n} \setminus c_{n+1},...,c_{n+m} \Leftrightarrow G | B \in \Gamma_{C}(r)}
        \hypo{\tau_{\tbool} \models G}
        \infer2[apply]{\left\{c_1, ..., c_{n+m}\right\} \uplus \Delta S \xmapsto{\left\{r\right\}} \left\{c_1, ..., c_n\right\} \uplus B \uplus \Delta S}
      \end{prooftree}
    \end{align*}
    \caption{\emph{Very abstract} operational semantics for ground and pure CHR}
    \label{fig-210-omega-a-semantics}
  \end{figure}
\end{definition}

Rules are applied until no more are applicable to the state, \IE we have reached a \emph{final} state.

\begin{example}[$\oma$-transitions]\label{ex::chr:sem:op:very-abstract}
  Intuitively, both
  \begin{align*}
    \tau_{\tbool} \models 0 < 4 \band 0 < 6 \band 4 \leq 6
    && \mbox{and} &&
    \tau_{\tbool} \models \bnot (0 < 0 \band 0 < 6 \band 0 \leq 6)
  \end{align*}
  hold.
  Hence, we can prove the transition $\left\{4, 6\right\} \xmapsto{\left\{subtract\right\}} \left\{4,2\right\}$, but not $\left\{0, 6\right\} \xmapsto{\left\{subtract\right\}} \left\{0,6\right\}$.
\end{example}

The following example shows the execution of the Euclidean algorithm as a final example of the operational semantics of CHR.

\begin{example}[Euclidean algorithm (cont.)]
  The rules of \textsc{gcd} are applied until exhaustion, leaving only the greatest common divisor of all numbers of the input.
  For an input $\left\{4,6\right\}$, the program will perform a sequence
  \begin{align*}
    \left\{4,6\right\}
    \xmapsto{\left\{subtract\right\}} \left\{4,2\right\}
    \xmapsto{\left\{subtract\right\}} \left\{2,2\right\}
    \xmapsto{\left\{subtract\right\}} \left\{2,0\right\}
    \xmapsto{\left\{zero\right\}} \left\{2\right\}
  \end{align*}
  of transformations.
\end{example}

  \section{FreeCHR}\label{sec::freechr:freechr}
The main idea of FreeCHR is to model the syntax of CHR programs as a functor $\chr_C$.
We then use the \emph{free} $\chr_C$-algebra to define the operational semantics of FreeCHR. 

\subsection{Syntax}\label{sec::freechr:fr:syntax}
We first present the fundamental definition of our work which allows us to model CHR-programs over a domain $C$.

\begin{definition}[Syntax of FreeCHR programs]\label{def::freechr:syn:set:functor}
  The functor\footnote{That $\chr_C$ is indeed a functor can easily be verified via equational reasoning.}
  \begin{align*}
    \chr_C D =\
      &\flist \tbool^C \times \flist \tbool^C \times \tbool^{\flist C} \times (\fmultiset C)^{\flist C} \sqcup D \times D\\
    \chr_C f =\
      &\id \sqcup f \times f
  \end{align*}
  describes the syntax of FreeCHR programs.
\end{definition}

The set $\flist\tbool^C \times \flist\tbool^C \times \tbool^{\flist C} \times (\fmultiset C)^{\flist C}$ is the set of single rules.
The kept and removed head of a rule are sequences of functions in $\flist \tbool^C$ which map elements of $C$ to Booleans, effectively checking individual values for applicability of the rule.
The guard of the rule is a function in $\tbool^{\flist C}$ and maps sequences of elements in $C$ to Booleans, checking all matched values in the context of each other.
Finally, the body of the rule is a function in $(\fmultiset C)^{\flist C}$ and maps the matched values to a multiset of newly generated values.

The set $D \times D$ represents the composition of FreeCHR programs by an execution strategy, allowing the construction of more complex programs from, ultimately, single rules.

By the structure of $\chr_C$, a $\chr_C$-algebra with carrier $D$ is defined by two functions 
\begin{align*}
  \rho :\ \flist \tbool^C \times \flist \tbool^C \times \tbool^{\flist C} \times (\fmultiset C)^{\flist C} \longrightarrow D &&
  \nu :\ D \times D \rightarrow D 
\end{align*}
as $(D, \left[\rho, \nu\right])$.
The free $\chr_C$-algebra $\chrstar$ provides us with an inductively defined representation of programs which we will later use to lift the very abstract operational semantics $\oma$.

\begin{lemma}[Free $\chr_C$-algebra]\label{lem::freechr:syn:set:free-algebra}
  With
  \begin{align*}
    \mu\chr_C 
      = \flist \tbool^C \times \flist \tbool^C \times \tbool^{\flist C} \times (\fmultiset C)^{\flist C} \sqcup \mu\chr_C \times \mu\chr_C
  \end{align*}
  and labels/injections
  \begin{align*}
    rule &: \flist \tbool^C \times \flist \tbool^C \times \tbool^{\flist C} \times (\fmultiset C)^{\flist C} \longrightarrow \mu\chr_C \\
    \chrcomp &: \mu\chr_C \times \mu\chr_C \longrightarrow \mu\chr_C
  \end{align*}
  $\chr^{\star}_C = (\mu\chr_C, \left[rule, \chrcomp\right])$ is the \emph{free} $\chr_C$-algebra.
\end{lemma}
\begin{proof}
  We show that $\chr^{\star}_C = (\mu\chr_C, \left[rule, \chrcomp\right])$ is the \emph{free} $\chr_C$-algebra, by constructing the $\chr_C$-catamorphism
  \begin{align*}
    \kata{\left[\rho,\nu\right]} : \mu\chr_C \longrightarrow \left(A, \left[\rho, \nu\right]\right)
  \end{align*}
  for any $\chr_C$-algebra $\left(A, \left[\rho, \nu\right]\right)$.
  By definition of the free $\chr_C$-algebra,
  \begin{align}
    \kata{\left[\rho,\nu\right]}(\left[rule, \chrcomp\right](p)) = \left[\rho, \nu\right]((\chr_C \kata{\left[\rho,\nu\right]})(p)) \label{eq::freechr:syn:set:fa}
  \end{align}
  needs to be true for any $p \in \mu\chr_C$.

  \begin{case*}[$p = rule(k, r, g, b)$]
    \begin{align*}
      &\ \kata{\left[\rho,\nu\right]}(\left[rule, \chrcomp\right](k,r,g,b)) = \left[\rho, \nu\right]((\chr_C \kata{\left[\rho,\nu\right]})(k,r,g,b)) \\ 
      \Leftrightarrow &\ \kata{\left[\rho,\nu\right]}(rule(k,r,g,b)) = \left[\rho, \nu\right](\id(k,r,g,b)) \\ 
      \Leftrightarrow &\ \kata{\left[\rho,\nu\right]}(rule(k,r,g,b)) = \left[\rho, \nu\right](k,r,g,b) \\ 
      \Leftrightarrow &\ \kata{\left[\rho,\nu\right]}(rule(k,r,g,b)) = \rho(k,r,g,b) \\ 
    \end{align*}
  \end{case*}
  \begin{case*}[$p = p_1 \chrcomp p_2$]
    \begin{align*}
      &\ \kata{\left[\rho,\nu\right]}(\left[rule, \chrcomp\right](p_1, p_2)) = \left[\rho, \nu\right]((\chr_C \kata{\left[\rho,\nu\right]})(p_1, p_2)) \\ 
      \Leftrightarrow &\ \kata{\left[\rho,\nu\right]}(p_1 \chrcomp p_2) = \left[\rho, \nu\right]((\kata{\left[\rho,\nu\right]} \times \kata{\left[\rho,\nu\right]})(p_1, p_2)) \\ 
      \Leftrightarrow &\ \kata{\left[\rho,\nu\right]}(p_1 \chrcomp p_2) = \left[\rho, \nu\right](\kata{\left[\rho,\nu\right]}(p_1), \kata{\left[\rho,\nu\right]}(p_2)) \\ 
      \Leftrightarrow &\ \kata{\left[\rho,\nu\right]}(p_1 \chrcomp p_2) = \nu(\kata{\left[\rho,\nu\right]}(p_1), \kata{\left[\rho,\nu\right]}(p_2)) \\ 
    \end{align*}
  \end{case*}
  Therefore, (\ref{eq::freechr:syn:set:fa}) holds \IFF
  \begin{align*}
    \kata{\left[\rho, \nu\right]} :&\ \mu\chr_C \longrightarrow A\\ 
    \kata{\left[\rho, \nu\right]}(rule(k,r,g,b)) =&\ \rho(k,r,g,b)\\
    \kata{\left[\rho, \nu\right]}(p_1 \chrcomp p_2) =&\ \nu(\kata{\left[\rho, \nu\right]}(p_1), \kata{\left[\rho, \nu\right]}(p_2))
  \end{align*}
  for all $\chr_C$-algebras $\left(A, \left[\rho, \nu\right]\right)$.
  Hence, there is a unique homomorphism 
  \begin{align*}
    \kata{\left[\rho, \nu\right]} : \chr^{\star}_C \rightarrow (A, \left[\rho, \nu\right])
  \end{align*}
  for any $\chr_C$-algebra $\left(A, \left[\rho, \nu\right]\right)$, making $\chr^{\star}_C$ the free $\chr_C$-algebra.
\end{proof}

The free $\chr_C$-algebra corresponds to the definition of abstract syntax trees of programs,
while the catamorphism $\kata{\alpha}$ corresponds to an interpretation that preserves the semantics of $\alpha$.

We can easily see that $\chrcomp$ is associative up to isomorphism\footnote{By $\assoc(a, (b, c)) = ((a, b), c)$ and $\assoc^{-1}((a, b), c) = (a, (b, c))$.}. We thus will not explicitly write parentheses and generally use chained expressions like $p_1 \chrcomp ... \chrcomp p_l$ for some $l \in \mathbb{N}$.

\begin{example}[Euclidean algorithm (cont.)]\label{ex::freechr:syn:set:fa:gcd}
  The program $\mathit{gcd} = \mathit{zero} \odot \mathit{subtract}$ with
  \begin{align*}
      \mathit{zero} &= rule(\eps, (\lambda n. n = 0), (\lambda n. \btrue), (\lambda n. \emptyset)) \\
      \mathit{subtract} &= rule((\lambda n. 0 < n), (\lambda m. 0 < m), (\lambda n\ m. n \leq m), (\lambda n\ m. \left\{m-n\right\})) 
  \end{align*}
  implements the euclidean algorithm, as defined in \autoref{ex::freechr:pre:chr:gcd}.
  $\lambda$-abstractions are used for ad-hoc definitions of functions.
\end{example}

\subsection{Operational semantics}\label{sec::freechr:fr:operational-semantics}
We now lift the \emph{very abstract} operational semantics $\omega_a$ of CHR to the very abstract operational semantics $\omstara$ of FreeCHR.
We assume that our programs are defined over a domain $C$, where there is a data type $\mathcal{L} C = \langle C, \Lambda_C, \tau_C \rangle$.
Like $\oma$, $\omstara$ is defined as a labelled transition system, where states are multisets over elements of $C$.

\begin{definition}[\emph{Very abstract} operational semantics $\omstara$]\label{def::freechr:sem:op:very-abstract}
  The \emph{very abstract} operational semantics of FreeCHR is defined as the labelled transition system
  \begin{align*}
    \omstara = \langle \fmultiset C, \mu\chr_C, \mapsto \rangle
  \end{align*}
  where the transition relation $\left(\mapsto\right)$ is defined by the inference rules in \autoref{fig::freechr:sem:op:very-abstract}.
  
  \begin{figure}[t]
    \begin{align*}
      \begin{prooftree}[small]
        \infer0[pass/final]{S \freetr{p} S}
      \end{prooftree}
      &&
      \begin{prooftree}[small]
        \hypo{S \freetr{p_i} S'}
        \hypo{S' \freetr{p_1 \chrcomp ... \chrcomp p_i \chrcomp ... \chrcomp p_l} S''}
        \infer2[step$_i$]{S \freetr{p_1 \chrcomp ... \chrcomp p_i \chrcomp ... \chrcomp p_l} S''}
      \end{prooftree}\\
    \end{align*}
    \begin{align*}
      \begin{prooftree}[small]
        \hypo{k_1(c_1) \band ... \band k_n(c_n) \band r_1(c_{n+1}) \band ... \band r_m(c_{n+m}) \band g(c_1,...,c_{n+m}) \equiv_{\tbool} \btrue}
        \infer1[apply]{\left\{c_1,  ...,  c_{n+m}\right\} \uplus \Delta S \freetr{rule(k, r, g, b) } \left\{c_1, ..., c_{n}\right\} \uplus b\left(c_1,...,c_{m+n}\right) \uplus \Delta S}
      \end{prooftree}
    \end{align*}
    \caption{\emph{Very abstract} operational semantics of FreeCHR}
    \label{fig::freechr:sem:op:very-abstract}
  \end{figure}
\end{definition}

The rule \textsc{pass/final} states that a program is always allowed to do nothing to a state.
\textsc{step$_i$} states that we can derive a transition from $S$ to $S''$, if we can transition from $S$ to $S'$ with the $i$-th program in the composition $p_1 \chrcomp ... \chrcomp p_l$ (for $1 \leq i \leq l$) and from $S'$ to $S''$ with the whole composition.
The idea is that, \WLOG $p_i$ is a rule that is applied to the current state, whereafter execution is continued.
\textsc{apply} is the translation of the original \textsc{apply}-rule of $\oma$.
$k_i$ and $r_j$ denote the $i$-th and $j$-th elements of the sequences $k$ and $r$.

We want to demonstrate the \emph{very abstract} operational semantics of FreeCHR on a small example.
\begin{example}[Euclidean algorithm (cont.)]\label{ex:freechr:gcd:exec}
  Given the program \emph{gcd} of \autoref{ex::freechr:syn:set:fa:gcd} and the initial state $\left\{12, 9\right\}$, we can prove the derivation
  \begin{align*}
    \left\{12, 9\right\} \xmapsto{gcd} \left\{3\right\}
  \end{align*}
  as follows.
  We first use the \textsc{step} transition to apply the \textit{subtract} rule to the initial state.
  \begin{align*}
    \begin{prooftree}[small]
      \hypo{\left\{12, 9\right\} \xmapsto{subtract} \left\{3, 9\right\}}
      \hypo{\left\{3, 9\right\} \xmapsto{zero \chrcomp subtract} \left\{3\right\}}
      \infer2[step$_2$]{\left\{12, 9\right\} \xmapsto{zero \chrcomp subtract} \left\{3\right\}}
    \end{prooftree}
  \end{align*}
  and prove that this transition is valid
  \begin{align*}
    \begin{prooftree}[small]
      \hypo{0 < 9 \wedge 0 < 12 \wedge 9 \leq 12 \equiv \btrue}
      \infer1[apply]{\left\{12, 9\right\} \uplus \emptyset \xmapsto{subtract} \left\{9\right\} \uplus \left\{3\right\} \uplus \emptyset}
    \end{prooftree}
  \end{align*}
  We repeat this process for the transitions $\left\{3, 9\right\} \xmapsto{gcd} \left\{3, 6\right\}$, $\left\{3, 6\right\} \xmapsto{gcd} \left\{3, 3\right\}$ and $\left\{3, 3\right\} \xmapsto{gcd} \left\{3, 0\right\}$.
  At this point, we have to apply the \textit{zero} rule to remove the $0$ and use the \textsc{pass/final} transition to prove the empty transition $\left\{3\right\} \xmapsto{gcd} \left\{3\right\}$.
  \begin{align*}
    \begin{prooftree}[small]
      \hypo{0 = 0 \wedge \btrue \equiv \btrue}
      \infer1[apply]{\left\{3, 0\right\} \uplus \emptyset \xmapsto{zero} \left\{3\right\} \uplus \emptyset \uplus \emptyset}
    \end{prooftree}
    &&
    \begin{prooftree}[small]
      \hypo{\left\{3, 0\right\} \xmapsto{zero} \left\{3\right\}}
      \infer0[pass/final]{\left\{3\right\} \xmapsto{zero \chrcomp subtract} \left\{3\right\}}
      \infer2[step$_2$]{\left\{3, 0\right\} \xmapsto{zero \chrcomp subtract} \left\{3\right\}}
    \end{prooftree}
  \end{align*}
  With this done, we have fully proven the derivation.
\end{example}

\subsection{Soundness and completeness of $\omstara$}\label{sec::freechr:fr:correctness}
To prove the soundness and completeness of $\omstara$ \WRT $\oma$, we first need to embed FreeCHR into the \emph{positive range-restricted ground} segment of CHR.
This is a common approach to relate rule-based formalisms to CHR \cite[Chapter 6]{fruehwirth2009constraint}.

\begin{definition}[Embedding FreeCHR into CHR]\label{def::freechr:syn:set:transformation-freechr-to-chr}
  Let
  \begin{align*}
    \fembed &: \mu \chr_C \longrightarrow \prgc\\
      \fembed\left(rule\left(k, r, g, b\right)\right)
      &=  \left\{R@v_1, ..., v_n \setminus v_{n+1}, ..., v_{n+m} \Leftrightarrow G | b\left(v_1, ..., v_{n+m}\right) \right\} \\
    \fembed\left(p_1 \chrcomp ... \chrcomp p_l\right) &= \fembed\left(p_1\right) \cup ... \cup \fembed\left(p_l\right) 
  \end{align*}
  be the function embedding FreeCHR programs into the \emph{positive range-restricted ground} segment of CHR, 
  with universally quantified variables $v_1$,...,$v_{n+m}$, $R$ a uniquely generated rule name and
  \begin{align*}
    G = k_1\left(v_1\right) \band ... \band r_m\left(v_{n+m}\right) \band g\left(v_1, ..., v_{n+m}\right)
  \end{align*}
\end{definition}

\begin{example}
  If we apply the embedding $\fembed$ to the program $gcd$ (including the evaluation of function applications\footnote{\IE $\beta$-reductions}), we get the program
  \begin{align*}
    zero\ &@\ \eps\ \setminus\ n\ \Leftrightarrow\ n = 0 \band \btrue\ |\ \emptyset \\
    subtract\ &@\ n\ \setminus\ m\ \Leftrightarrow\ 0 < n \band 0 < m \band n \leq m\ |\ m - n
  \end{align*}
  with universally quantified variables $n$ and $m$.
  The program corresponds to the \emph{head-normalization} (\cite{duck2005compilation}) of the program in \autoref{ex::freechr:pre:chr:gcd}.
\end{example}

We now prove $\fembed$-soundness of $\omstara$ \WRT $\oma$, \IE if we can prove a derivation under $\omstara$ for a program $p$, we can prove it under $\oma$ for $\fembed(p)$.

\begin{theorem}[$\fembed$-Soundness of $\omstara$]\label{th::freechr:sem:op:set:va:soundness}
  $\omstara$ is \emph{$\fembed$-sound} \WRT $\oma$, \IE for $S, S' \in \fmultiset C$ and $p \in \mu\chr_C$,
  \begin{align*}
    S \freetr{p} S' \Longrightarrow S \freetr{\fembed\left(p\right)} S'
  \end{align*}
\end{theorem} 
\begin{proof}
  We prove soundness by induction over the inference rules of $\omstara$.
  \begin{indbase}[\textsc{pass/final}]
    Follows directly from \textsc{empty}.
  \end{indbase}

  \begin{indbase}[\textsc{apply}]
    Given a proof
    \begin{align*}
      \begin{prooftree}[small]
        \hypo{G \equiv_{\tbool} \btrue}
        \infer1[apply]{
          \left\{c_1, ..., c_{n+m}\right\} \uplus \Delta S
          \freetr{p} \left\{c_1, ..., c_n\right\} \uplus b(c_1, ..., c_{n+m}) \uplus \Delta S
        }
      \end{prooftree}
    \end{align*}
    with $r = rule(\left(k_1, ..., k_n\right), \left(r_1, ..., r_m\right), g, b)$ and
    \begin{align*}
      G = k_1\left(t_1\right) \band ... \band k_n\left(t_n\right) \band r_1\left(t_{n+1}\right) \band ... \band r_m\left(t_{n+m}\right) \band g\left(t_1, ..., t_{n+m}\right)
    \end{align*}
    there is a $C$-instance
    \begin{align*}
      R\ @\ c_1, ..., c_n\ \setminus\ c_{n+1}, ..., c_{n+m}\ \Leftrightarrow\ G\ |\ b(c_1, ..., c_{n+m}) \in \Gamma_C(\fembed(p))
    \end{align*}
    such that $\tau_{\tbool} \models G$.
    By \autoref{def::freechr:syn:set:transformation-freechr-to-chr} we know that $\left\{p'\right\} = \fembed\left(p\right)$.
    We can thus prove $S \freetr{\fembed(p)} S'$ by
    \begin{align*}
      &\begin{prooftree}[small]
        \hypo{R @ c_1, ..., c_n \setminus c_{n+1}, ..., c_{n+m} \Leftrightarrow G | b(c_1, ..., c_{n+m}) \in \Gamma_C\left(\fembed(p')\right)}
        \hypo{\tau_{\tbool} \models G}
        \infer2[apply]{
          \underbrace{\left\{c_1, ..., c_{n+m}\right\} \uplus \Delta S}_{= S}
          \freetr{\{p'\}}
          \underbrace{\left\{c_1, ..., c_n\right\} \uplus b(c_1, ..., c_{n+m}) \uplus \Delta S}_{= S'}}
      \end{prooftree}
    \end{align*}
  \end{indbase}

  \begin{indstep}[\textsc{step}$_i$]
    Given a proof
    \begin{align*}
      \begin{prooftree}[small]
        \hypo{S \freetr{p_i} S'}
        \hypo{S' \freetr{p} S''}
        \infer2[step$_i$]{S \freetr{p} S''}
      \end{prooftree}
    \end{align*}
    with $p = p_1 \chrcomp ... \chrcomp p_i \chrcomp ...  \chrcomp p_l$, for $1 \leq i \leq l$, we assume
    \begin{align}
      \forall i \in {1,...,l}. S \freetr{p_i} S' \Longrightarrow S \freetr{\fembed(p_i)} S' \label{eq::freechr:sem:op:set:va:soundness:1}\\
      S' \freetr{p} S'' \Longrightarrow S' \freetr{\fembed(p)} S''\label{eq::freechr:sem:op:set:va:soundness:2}
    \end{align}
    as induction hypotheses.

    Since we can, \WLOG, assume that $p_i = \chrrule(...)$, we know from \autoref{def::freechr:syn:set:transformation-freechr-to-chr} that $\left\{p'\right\} = \fembed(p_i) \subseteq \fembed(p)$.
    By induction hypothesis (\ref{eq::freechr:sem:op:set:va:soundness:1}) we know from $S \freetr{p_i} S'$ that $S \freetr{\fembed(p_i)} S'$.
    By induction hypothesis (\ref{eq::freechr:sem:op:set:va:soundness:2}) we know from $S' \freetr{p} S''$ that $S' \freetr{\fembed(p)} S''$.
    We can thus prove $S \freetr{\fembed(p)} S''$ by
    \begin{align*}
      \begin{prooftree}[small]
        \hypo{p' \in \fembed(p)}
        \hypo{S \freetr{\left\{p'\right\}} S'}
        \hypo{S' \freetr{\fembed(p)} S''}
        \infer3[step]{S \freetr{\fembed(p)} S''}
      \end{prooftree}
    \end{align*}
  \end{indstep}
\end{proof}

We established that we can prove any derivation we can prove with a program $p$ under $\omstara$, with a program $\fembed(p)$ under $\oma$.
We also want to prove $\fembed$-completeness, \IE we can prove any derivation with a program $\fembed(p)$ under $\oma$ with a program $p$ under $\omstara$.

\begin{theorem}[$\fembed$-completeness of $\omstara$]\label{th::freechr:sem:op:set:va:completeness}
  $\omstara$ is \emph{$\fembed$-complete} \WRT $\oma$,
  \IE for $S, S' \in \fmultiset C$ and $p \in \mu\chr_C$,
  \begin{align*}
     S \freetr{\fembed\left(p\right)} S' \Longrightarrow S \freetr{p} S'
  \end{align*}
\end{theorem}
\begin{proof}
  We prove $\fembed$-completeness by induction over the inference rules of $\oma$

  \begin{indbase}[\textsc{empty}]
    Follows directly from \textsc{pass/final}
  \end{indbase}

  \begin{indbase}[\textsc{apply}]
    Given a proof
    \begin{align*}
      \begin{prooftree}[small]
        \hypo{R@c_{1},...,c_{n} \setminus c_{n+1},...,c_{n+m} \Leftrightarrow G | B \in \Gamma_{C}(\fembed(p))}
        \hypo{\tau_{\tbool} \models G}
        \infer2[apply]{\left\{c_1, ..., c_{n+m}\right\} \uplus \Delta S \freetr{\fembed(p)} \left\{c_1, ..., c_n\right\} \uplus B \uplus \Delta S}
      \end{prooftree}
    \end{align*}

    By definition of $\oma$, we know that $\fembed(p) = \left\{p'\right\}$.
    Therefore, $p = rule(k, r, g, b)$, and hence $G = k_1(c_1) \band ... \band r_m(c_{n+m}) \band g(c_1,...,c_{n+m})$ and $B = b(c_1,...,c_{n+m})$.
    From $\tau_{\tbool} \models G$ follows $G \equiv_{\tbool} \btrue$.
    We can thus construct a proof
    \begin{align*}
      \begin{prooftree}[small]
        \hypo{G \equiv_{\tbool} \btrue}
        \infer1[apply]{\left\{c_1, ..., c_{n+m}\right\} \uplus \Delta S \freetr{p} \left\{c_1, ..., c_n\right\} \uplus B \uplus \Delta S}
      \end{prooftree}
    \end{align*}
  \end{indbase}

  \begin{indstep}[\textsc{step}]
    Given a proof
    \begin{align*}
      \begin{prooftree}[small]
        \hypo{p' \in \fembed(p)}
        \hypo{S \freetr{\left\{r'\right\}} S'}
        \hypo{S' \freetr{\fembed(p)} S''}
        \infer3[step]{S \freetr{\fembed(p)} S''}
      \end{prooftree}
    \end{align*}
    By \autoref{def::freechr:syn:set:transformation-freechr-to-chr}, we know from $r' \in \fembed(p)$ that for $p = p_1 \chrcomp ... \chrcomp p_i \chrcomp ... \chrcomp p_l$ (with $l \geq 1$) there is a $p_i$ such that $\left\{p'\right\} = \fembed(p_i)$.
    We hence assume as induction hypotheses that
    \begin{align}
      S \freetr{\left\{p'\right\}} S' \Longrightarrow S \freetr{p_i} S' \label{eq::completeness:omstara:1}\\
      S' \freetr{\fembed(p)} S'' \Longrightarrow S' \freetr{p} S'' \label{eq::completeness:omstara:2}
    \end{align}
    From $S \freetr{\left\{r\right\}} S'$ and induction hypothesis (\ref{eq::completeness:omstara:1}) follows $S \freetr{p_i} S'$.
    From $S' \freetr{\fembed(p)} S''$ and induction hypothesis (\ref{eq::completeness:omstara:2}) follows $S' \freetr{p} S''$.
    We can hence construct the proof
    \begin{align*}
      \begin{prooftree}[small]
        \hypo{S \freetr{p_i} S'}
        \hypo{S' \freetr{p} S''}
        \infer2[step]{S \freetr{p} S''}
      \end{prooftree}
    \end{align*}
  \end{indstep}
\end{proof}

With \autoref{th::freechr:sem:op:set:va:soundness} and \autoref{th::freechr:sem:op:set:va:completeness}, we have established that, up to $\fembed$, FreeCHR is as expressive as CHR.
Analogously to classical CHR, we are now able to define more refined operational semantics for FreeCHR and prove their soundness \WRT $\omstara$, which is ongoing future work.

  \section{An instance with \emph{very abstract} operational semantics}\label{sec::freechr:instance:very-abstract}
In this section, we want to present a simple execution algorithm and prove its adherence to $\omstara$ by using the inductive nature of our definition of syntax and semantics of FreeCHR.
This will serve both as a first, albeit simple, guideline for implementations, as well as a case study on how a FreeCHR execution algorithm can be defined and verified.

\subsection{Instances}\label{sec::freechr:instance}

Since the operational semantics define the execution of FreeCHR programs as a state transition system, an obvious way to implement FreeCHR is by mapping programs to functions mapping one state to its successor.

\begin{definition}[FreeCHR instance]
  An \emph{instance} of FreeCHR is a $\chr_C$-algebra
  \begin{align*}
    \left((\fmultiset C)^{\fmultiset C}, \left[\chrinstrule, \chrinstcomp\right] \right)
  \end{align*}
  with
  \begin{align*}
    \chrinstrule &: \flist \tbool^C \setprod \flist \tbool^C \setprod \tbool^{\flist C} \setprod \left(\fmultiset C\right)^{\flist C} \longrightarrow (\fmultiset C)^{\fmultiset C}\\
    \chrinstcomp &: (\fmultiset C)^{\fmultiset C} \setprod (\fmultiset C)^{\fmultiset C} \longrightarrow (\fmultiset C)^{\fmultiset C}
  \end{align*}
\end{definition}

A FreeCHR instance hence needs to define the two functions $\chrinstrule$ and $\chrinstcomp$ which implement the execution of programs and hence the operational semantics of the program.
$\chrinstcomp$ implements the application of a single rule to a state if it is possible. $\chrinstcomp$ implements the execution strategy of, fundamentally, a sequence of rules to a state.
Both are only concerned with single execution steps.
We hence also need the function
\begin{align*}
  \mathtt{run} : (\fmultiset C)^{\fmultiset C} \longrightarrow (\fmultiset C)^{\fmultiset C}
\end{align*}
which maps a function performing singular steps to a function applying this function until a final state is reached. In CHR this generally means, until no rules are applicable to the state.

Since an instance practically also defines operational semantics it makes sense to directly associate it with an LTS as well.
\begin{definition}[LTS of a FreeCHR instance]\label{def::freechr:instance-rel}
  Let
  \begin{align*}
    I = \left((\fmultiset C)^{\fmultiset C}, \left[\rho, \nu\right]\right) 
  \end{align*}
  $I$ implies an LTS
  \begin{align*}
    \trrel I = \langle \fmultiset C, L, \mapsto\rangle
  \end{align*}
  where the labels $L$ are recursively defined as
  \begin{align*}
    L =& \left\{\rho(k, r, g, b) \mid k, r \in \flist (\tbool^C), g \in \tbool^{\flist C}, b \in (\fmultiset C)^{\flist C}\right\}
      \cup \left\{\nu(f, g) \mid f, g \in L \right\}
  \end{align*}
  and the transition relation as
  \begin{align*}
    (\mapsto) =& \left\{s \xmapsto{f} s' \mid s \in \fmultiset C, f \in L, f(s) \equiv_{\fmultiset C} s'\right\}
  \end{align*}
\end{definition}

With this, we can easily verify an instance against operational semantics, using the concepts from \autoref{sec::lts}.
We will do so from the perspective of the operational semantics, instead of from the perspective of the implementation.
To verify, that an instance only performs valid transitions, we hence prove \emph{$\fembed$-completeness}
\begin{align*}
  S \freetr{p} S' \Longleftarrow S \freetr{\kata{\alpha}(p)} S'
\end{align*}
Consequentially, to prove that the implementation can perform all valid transitions, we prove \emph{$\fembed$-soundness up to repeated transition}
\begin{align*}
  S \freetr{p} S' \Longrightarrow S \xmapsto{\kata{\alpha}(p)}^{*} S'
\end{align*}
We need the relaxation \emph{"up to repeated transition"} since the instances, and hence their respective LTS, are only defined for singular steps.

\subsection{Execution algorithm}\label{sec::freechr:instance:very-abstract:alg}
The algorithm presented in this section implements the operations required for a FreeCHR instance as non-deterministic imperative pseudo-code.
The $\chrrule$ function is implemented by $\textsc{rule}$, $\chrcomp$ by $\textsc{compose}$, in \autoref{alg::freechr:imp:set:alg:very-abstract}.

\begin{definition}[Pure CHR-programs in \textbf{Set} (cont.)]\label{def::freechr:imp:set:alg:very-abstract}
  \autoref{alg::freechr:imp:set:alg:very-abstract} shows the abstract implementation of a $\chr_C$-algebra
  \begin{align*}
    \ExecA = \left((\fmultiset C)^{\fmultiset C}, \left[\textsc{rule}, \textsc{compose}\right]\right)
  \end{align*}

  \begin{algorithm}[t]
    {\small
    \begin{algorithmic}[1]
      \Function{rule}{$\chrkept$, $\chrremoved$, $\chrguard$, $\chrbody$, $S$}
        \If{$\exists \left\{c_1, ..., c_{n+m}\right\} \subseteq S. \chrkept_1(c_1) \band ... \band \chrremoved_m(c_{n+m}) \band \chrguard(c_1,...,c_{n+m}) \equiv_{\tbool} \btrue$}
          \label{alg::freechr:imp:set:alg:very-abstract:1}
          \State \Return $\left(S \msetminus \left\{c_{n+1}, ..., c_{n+m}\right\}\right) \uplus \chrbody(c_1,...,c_{n+m})$
          \label{alg::freechr:imp:set:alg:very-abstract:2}
        \Else
          \State \Return $S$
          \label{alg::freechr:imp:set:alg:very-abstract:3}
        \EndIf
      \EndFunction
      \\
      \Function{compose}{$(p_1,...,p_l)$, $S$}
        \If{$\exists p \in \left\{p_1,..., p_l\right\}. p(S) \not\equiv_{\fmultiset C} S$}
          \label{alg::freechr:imp:set:alg:very-abstract:4}
          \State \Return $p(S)$
          \label{alg::freechr:imp:set:alg:very-abstract:5}
        \Else
          \State \Return $S$
          \label{alg::freechr:imp:set:alg:very-abstract:6}
        \EndIf
      \EndFunction
    \end{algorithmic}}
    \caption{Implementation of $\ExecA$}
    \label{alg::freechr:imp:set:alg:very-abstract}
  \end{algorithm}
\end{definition}

The function \textsc{rule} first checks whether there is a matching $\left\{c_1, ..., c_{n+m}\right\} \subseteq S$, such that head constraints $\chrkept_i$ applied to $c_i$ and $\chrremoved_j$ applied to $c_{n+j}$ evaluate to $\btrue$, as well as the guard, applied to all constraints of the matching (l. \ref{alg::freechr:imp:set:alg:very-abstract:1}).
If such a subset exists, the constraints $c_{n+j}$ with $1 \leq j \leq m$ are removed and the multiset resulting from applying the body to the matching will be added to $S$ (l. \ref{alg::freechr:imp:set:alg:very-abstract:2}).
The function \textsc{compose} checks if there is a $p\in\left\{p_1,...,p_l\right\}$ with $p(S) \neq S$ (l. \ref{alg::freechr:imp:set:alg:very-abstract:4}).
If this is the case, $p$ is applied to $S$ (l. \ref{alg::freechr:imp:set:alg:very-abstract:5}).
If in either function the respective check does not succeed, $S$ is returned unmodified (l. \ref{alg::freechr:imp:set:alg:very-abstract:3} and \ref{alg::freechr:imp:set:alg:very-abstract:6}, respectively). The execution strategy implemented by \textsc{compose} is to non-deterministically select an applicable program and apply it to the state.

Note, that we will freely use currying on the last argument of either function, \FI $\textsc{rule}(k,r,g,b)$ is a function from $\fmultiset C$ to $\fmultiset C$, and $\textsc{rule}(k,r,g,b,S)$ evaluates to an $S' \in \fmultiset C$.
Also, $\textsc{compose}$ will accept an arbitrary number of subprograms (but always an input state).

Finally, we need to implement the $run$ function.

\begin{definition}[Pure CHR-programs in \textbf{Set} (cont.)]\label{def::freechr:imp:set:alg:run}
  \autoref{alg::freechr:imp:set:alg:run} implements application until exhaustion for FreeCHR programs defined using \autoref{alg::freechr:imp:set:alg:very-abstract}.
 \begin{algorithm}[t]
    {\small
    \begin{algorithmic}[1]
      \Function{run}{$f$, $S$}
        \Repeat
          \State $S' \leftarrow S$
          \State $S \leftarrow f(S)$
        \Until{$S = S'$}
        \State \Return $S$
      \EndFunction
    \end{algorithmic}}
    \caption{Implementation of $run_{f}$}
    \label{alg::freechr:imp:set:alg:run}
  \end{algorithm}
\end{definition}

$\textsc{run}(f,S)$ applies the function $f$ to the argument state $S$ repeatedly.
After each iteration, it is compared to the last state.
As soon as both are equal it is assumed that no more rules could be applied.
Then, the state is returned as the final state.

Note, that this algorithm would also terminate if a propagation rule with an empty body is applied.
As such rules are nonsensical, we will ignore this case.

\subsection{Correctness}\label{sec::freechr:fr:correctness-alg-1}
We will now prove the correctness of $\ExecA$.
With $\execA = \left[\textsc{rule},\textsc{compose}\right]$, let
\begin{align*}
  \kata{\execA} :\ &\chrstar \rightarrow \ExecA
\end{align*}
be the $\chr_C$-catamorphism to $\ExecA$, \IE the interpretation of programs in $\mu\chr_C$ that respects the semantics of \autoref{alg::freechr:imp:set:alg:very-abstract}.

\begin{theorem}\label{lem::freechr:imp:set:alg:ver:completeness}
  The \emph{very abstract} operational semantics $\omstara$ is \emph{$\kata{\execA}$-complete} \WRT \ExecA (\autoref{alg::freechr:imp:set:alg:very-abstract}), \IE
  \begin{align*}
    S \freetr{\kata{\execA}(p)} S' \Longrightarrow S \freetr{p} S'
  \end{align*}
\end{theorem}
\begin{proof}
  To prove the theorem, we will first discern two basic cases.
  \begin{case*}[$S = S'$]
    This case is trivial by rule \textsc{pass/final}.
  \end{case*}
  \begin{case*}[$S \neq S'$]
    We prove this case via induction over the structure of $p \in \chr_C$.
    \begin{indbase}[$p = \chrrule(k, r, g, b)$]
      The transition
      \begin{align*}
        S \freetr{\kata{\execA}(\chrrule(k, r, g, b))} S' 
      \end{align*}
      implies that
      \begin{align*}
        &\kata{\execA}(\chrrule(k, r, g, b))(S)\\
        =&\ \textsc{rule}(k, r, g, b, S)\\
        \equiv&\ S'
      \end{align*}
      From \autoref{alg::freechr:imp:set:alg:very-abstract} follows that the condition in line \ref{alg::freechr:imp:set:alg:very-abstract:1} must be true,
      since there were changes to the state.
      This implies that there is a subset $\left\{c_1, ..., c_{n+m}\right\} \subseteq S$ for which 
      \begin{align*}
        k_1(c_1) \wedge ... \wedge k_n(c_n) \wedge r_1(c_{n+1}) \wedge ... \wedge r_m(c_{n+m}) \wedge g(c_1,...,c_{n+m}) \equiv_{\tbool} \btrue
      \end{align*}
      Hence 
      \begin{align*}
        S &= \left\{c_1, ..., c_{n+m}\right\} \msetplus \Delta S
      \end{align*}
      and
      \begin{align*}
        (S \msetminus \left\{c_{n+1},...,c_{n+m}\right\}) \uplus b(c_1, ..., c_{n+m}) &= \left\{c_1, ..., c_n\right\} \msetplus b(c_1,...,c_n) \msetplus \Delta S
      \end{align*}
      From this, we can construct a proof
      \begin{align*}
        \begin{prooftree}[small]
          \hypo{k_1(c_1) \wedge ... \wedge k_n(c_n) \wedge r_1(c_{n+1}) \wedge ... \wedge r_m(c_{n+m}) \wedge g(c_1,...,c_{n+m}) \equiv_{\tbool} \btrue}
          \infer1[apply]{\left\{c_1, ..., c_{n+m}\right\} \msetplus \Delta S \freetr{\chrrule(k, r, g, b)} \left\{c_1, ..., c_n\right\} \msetplus b(c_1,...,c_n) \msetplus \Delta S}
        \end{prooftree}
      \end{align*}
    \end{indbase}
    \begin{indstep}[$p = p_1 \chrcomp ... \chrcomp p_n$]
      As the induction hypothesis, we assume
      \begin{align*}
        \forall p \in \chr_C. S \freetr{\kata{\execA}(p)} S' \Longrightarrow S \freetr{p} S'
      \end{align*}
      The transition
      \begin{align*}
        S \freetr{\kata{\execA}(p_1 \chrcomp ... \chrcomp p_n)} S'
      \end{align*}
      implies that
      \begin{align*}
        \kata{\execA}(p_1 \chrcomp ... \chrcomp p_n)(S)
        =\textsc{compose}(p_1, ...,p_n, S)
        \equiv_{\fmultiset C} S'
      \end{align*}
      From \autoref{alg::freechr:imp:set:alg:very-abstract}, especially line \ref{alg::freechr:imp:set:alg:very-abstract:4}, we know that there must hence be a $p_i$ with $1 \leq i \leq n$ such that $\kata{\execA}(p_i)(S) \equiv_{\fmultiset C} S'$, or in terms of the transition relation
      \begin{align*}
        S \freetr{\kata{\execA}(p_i)} S'
      \end{align*}
      By induction hypothesis, we then have a proof for $S \freetr{p_i} S'$.
      We can thus construct a proof
      \begin{align*}
        \begin{prooftree}[small]
          \hypo{S \freetr{p_i} S'}
          \infer0[pass/final]{S' \freetr{p_1 \chrcomp ... \chrcomp p_n} S'}
          \infer2[step$_i$]{S \freetr{p_1 \chrcomp ... \chrcomp p_n} S'}
        \end{prooftree}
      \end{align*}      
    \end{indstep}
  \end{case*}
\end{proof}

\begin{theorem}\label{lem::freechr:imp:set:alg:ver:soundness}
  The \emph{very abstract} operational semantics $\omstara$ is \emph{$\kata{\execA}$-sound up to repeated transition} \WRT \ExecA (\autoref{alg::freechr:imp:set:alg:very-abstract}), \IE
  \begin{align*}
    S \freetr{\kata{\execA}(p)}^{*} S' \Longleftarrow S \freetr{p} S'
  \end{align*}
\end{theorem}
\begin{proof}
  We prove the theorem by induction over the inference rules of $\omstara$.
  \begin{indbase}[\textsc{pass/final}]
    Trivial, since $S \freetr{\kata{\execA}(p)}^{*} S$ is always true for zero transition steps
  \end{indbase}
  \begin{indbase}[\textsc{apply}]
    Given a proof
    \begin{align*}
      \begin{prooftree}[small]
        \hypo{k_1(c_1) \band ... \band k_n(c_n) \band r_1(c_{n+1}) \band ... \band r_m(c_{n+m}) \band g(c_1,...,c_{n+m}) \equiv_{\tbool} \btrue}
        \infer1[apply]{\left\{c_1,...,c_{n+m}\right\} \msetplus \Delta S
          \freetr{\chrrule(k, r, g, b)} \left\{c_1,...,c_n\right\} \msetplus b\left(c_1, ..., c_{n+m}\right) \msetplus \Delta S}
      \end{prooftree}
    \end{align*}
    We want to show that
    \begin{align*}
      \left\{c_1,...,c_{n+m}\right\} \msetplus \Delta S \freetr{\kata{\execA}(\chrrule(k, r, g, b))} \left\{c_1,...,c_n\right\} \msetplus b\left(c_1, ..., c_{n+m}\right) \msetplus \Delta S
    \end{align*}
    or respectively
    \begin{align*}
      &\kata{\execA}(\chrrule(k, r, g, b))(\left\{c_1,...,c_{n+m}\right\} \msetplus \Delta S)\\
      =&\ \textsc{rule}(k, r, g, b, \left\{c_1,...,c_{n+m}\right\} \msetplus \Delta S) \\
      =&\ \left\{c_1,...,c_n\right\} \msetplus b\left(c_1, ..., c_{n+m}\right) \msetplus \Delta S
    \end{align*}
    Since by assumption the condition in line \ref{alg::freechr:imp:set:alg:very-abstract:1} will be $\btrue$.
    From line \ref{alg::freechr:imp:set:alg:very-abstract:2} we then know that 
    \begin{align*}
      &\textsc{rule}(k, r, g, b, \left\{c_1,...,c_{n+m}\right\} \msetplus \Delta S)\\
      \equiv& \left(\left\{c_1,...,c_{n+m}\right\} \msetplus \Delta S \msetminus \left\{c_{n+1}, ..., c_{n+m}\right\}\right) \msetplus b(c_1,...,c_{n+m})
    \end{align*}
    which can be simplified to $\left\{c_1,...,c_n\right\} \msetplus b\left(c_1, ..., c_{n+m}\right) \msetplus \Delta S$.
  \end{indbase}
  \begin{indstep}[\textsc{step}$_i$]
    Given a proof
    \begin{align*}
      \begin{prooftree}[small]
        \hypo{S \freetr{p_i} S'}
        \hypo{S' \freetr{p_1 \chrcomp ... \chrcomp p_l} S''}
        \infer2[step$_i$]{S \freetr{p_1 \chrcomp ... \chrcomp p_l} S''}
      \end{prooftree}
    \end{align*}
    with $1 \leq i \leq l$.
    We assume
    \begin{align}
      S \freetr{\kata{\execA}(p_i)}^{*} S' &\Longleftarrow S \freetr{p_i} S' \label{eq::freechr:imp:set:alg:ver:completeness:indhypo:1}\\
      S' \freetr{\kata{\execA}(p_1 \chrcomp ... \chrcomp p_l)}^{*} S'' &\Longleftarrow S' \freetr{p_1 \chrcomp ... \chrcomp p_l} S'' \label{eq::freechr:imp:set:alg:ver:completeness:indhypo:2}
    \end{align}
    as induction hypotheses.
    We want to show that
    \begin{align*}
      S \freetr{\kata{\execA}(p_1 \chrcomp ... \chrcomp p_l)}^{*} S''
    \end{align*}
    
    From the premises of the given proof we can infer
    $S \freetr{\kata{\execA}(p_i)}^{*} S'$ and $S' \freetr{\kata{\execA}(p_1 \chrcomp ... \chrcomp p_l)}^{*} S''$ by induction hypotheses (\ref{eq::freechr:imp:set:alg:ver:completeness:indhypo:1}) and (\ref{eq::freechr:imp:set:alg:ver:completeness:indhypo:2}), respectively.
    By the definition of \textsc{compose} we know that, since $\kata{\execA}(p_i)(S) \equiv S'$, $S'$ is also a possible outcome of $\kata{\execA}(p_1 \chrcomp ... \chrcomp p_l)(S)$.
    From $S \freetr{\kata{\execA}(p_i)}^{*} S'$, we can hence assume $S \freetr{\kata{\execA}(p_1 \chrcomp ... \chrcomp p_l)}^{*} S'$.
    Therefrom $S \freetr{\kata{\execA}(p_1 \chrcomp ... \chrcomp p_l)}^{*} S''$ follows directly.
  \end{indstep}
\end{proof}

With \autoref{lem::freechr:imp:set:alg:ver:completeness} and \autoref{lem::freechr:imp:set:alg:ver:soundness} we established that \autoref{alg::freechr:imp:set:alg:very-abstract} correctly implements the \emph{very abstract operational semantics} $\omstara$.
The theorems also imply that $\ExecA$ is an equivalent representation of $\omstara$.

\newcommand\snippetHSsolverStep[1][]{\inputminted[firstline=13,lastline=13,#1]{haskell}{code/FreeCHR.hs}}
\newcommand\snippetHSmatch[1][]{\inputminted[firstline=17,lastline=22,#1]{haskell}{code/FreeCHR.hs}}
\newcommand\snippetHSrule[1][]{\inputminted[firstline=26,lastline=34,#1]{haskell}{code/FreeCHR.hs}}
\newcommand\snippetHScompose[1][]{\inputminted[firstline=38,lastline=46,#1]{haskell}{code/FreeCHR.hs}}
\newcommand\snippetHSrun[1][]{\inputminted[firstline=53,lastline=58,#1]{haskell}{code/FreeCHR.hs}}
\newcommand\snippetHSgcd[1][]{\inputminted[firstline=62,lastline=68,#1]{haskell}{code/FreeCHR.hs}}
\newcommand\snippetHSfib[1][]{\inputminted[firstline=75,lastline=76,#1]{haskell}{code/FreeCHR.hs}}
\newcommand\snippetHSnub[1][]{\inputminted[firstline=87,lastline=88,#1]{haskell}{code/FreeCHR.hs}}
\newcommand\snippetHSalldiff[1][]{\inputminted[firstline=142,lastline=144,#1]{haskell}{code/FreeCHR.hs}}

\newcommand\snippetPYmatch[1][]{\inputminted[firstline=6,lastline=10,#1]{python}{code/FreeCHR.py}}
\newcommand\snippetPYrule[1][]{\inputminted[firstline=14,lastline=25,#1]{python}{code/FreeCHR.py}}
\newcommand\snippetPYcompose[1][]{\inputminted[firstline=29,lastline=37,#1]{python}{code/FreeCHR.py}}
\newcommand\snippetPYrun[1][]{\inputminted[firstline=41,lastline=51,#1]{python}{code/FreeCHR.py}}
\newcommand\snippetPYgcd[1][]{\inputminted[firstline=55,lastline=62,#1]{python}{code/FreeCHR.py}}
\newcommand\snippetPYfib[1][]{\inputminted[firstline=66,lastline=69,#1]{python}{code/FreeCHR.py}}
\newcommand\snippetPYnub[1][]{\inputminted[firstline=73,lastline=73,#1]{python}{code/FreeCHR.py}}
\newcommand\snippetPYalldiff[1][]{\inputminted[firstline=77,lastline=84,#1]{python}{code/FreeCHR.py}}

\section{Implementation case studies}\label{sec::freechr:instance:very-abstract:case-studies}
We now present two implementations that are based on the execution algorithm described in \autoref{sec::freechr:instance:very-abstract} as case studies.
The source code of the implementations is also available on GitHub\footnote{\url{https://gist.github.com/SRechenberger/d5e1eb875ae72ce5cafe6ea1c5b3ee38}}.

The purpose of this section is twofold:
first, we show how the abstract execution algorithm can be used as a guide to implement FreeCHR.
Second, we show how the chosen host language flavors the implementation of FreeCHR both in the back end, as well as in the front end.

The presented implementations have two limitations to be noted:
first, they solve any source of non-determinism in the original algorithm.
Whenever the original algorithm \emph{chooses} an element from a (multi) set, the implementations will sequentially search for them.
Especially is the state, in both implementations, a list instead of a multiset.
Also, the rules of a program will be traversed sequentially to find an applicable one.
It is, though, easy to see that they will only perform transitions \autoref{alg::freechr:imp:set:alg:very-abstract} would perform as well.
They are, in other words, concretizations of $\ExecA$.

Second, as they are based on the \emph{very abstract} operational semantics, they do not yet account for trivial non-termination due to repeated application of propagation rules (\IE rules with an empty removed head) (\cite{fruehwirth2009constraint}).
Hence, a program with such rules will not terminate if any of them is at some point applicable.
This problem is solved with the \emph{abstract} operational semantics by recording a \emph{propagation history}.
It saves applied constraints (by a unique identifier) as well as the rule applied to them.
If a propagation rule is to be applied, it is first checked if the combination was already executed.
If this is the case, the rule will not fire again.

The consequence for the presented implementations is that they only work on a subset of CHR/FreeCHR without propagation rules.

\subsection{Host Languages}
To sufficiently demonstrate how FreeCHR can be implemented in different programming languages and how the respective host language flavors the implementation,
we chose two programming languages of very different paradigms: \emph{Haskell} and \emph{Python}.

Haskell is a statically typed functional programming language with a powerful type system.
Python is a dynamically typed imperative language with plenty of practical applications.

One might consider both languages to be on the opposite of a static-dynamic/functional-imperative spectrum as well.
They will hence serve well for this case study, as most languages are featurewise somewhere in between.
Languages like Java or C\# successively adopt functional features but stay imperative in their core.
They also have a static type system like Haskell, but not as powerful and strict.
Languages like Kotlin move further towards the functional end of the spectrum but still inherit many characteristics from Java.
On the other hand, the whole LISP family of programming languages is functional but dynamically typed.

\subsection{Execution Algorithm}
We will now give a detailed overview of the implementations, connect them to the abstract algorithm and compare them to each other.

\subsubsection{Carrier}
We first want to discuss the implementation of the carrier of the implemented $\chr_C$-algebra.

In Haskell, the carrier is explicitly defined as the type \haskell{SolverStep c} shown in \autoref{fig::impl:solver:hs}.
The type is essentially equal to functions of type \haskell{[c] -> [c]} which is the Haskell representation of the set $(\fmultiset C)^{\fmultiset C}$.
As mentioned above, we will use lists instead of multisets.
This eliminates one source of non-determinism.

Wrapping the functions by defining a new type allows the implementation of type class instances (which is used for \textsc{compose}) and to discern programs that are "in construction" from such that were already passed to \haskell{run}.

The Python implementation directly models functions mapping lists to lists.
Since Python is (by default) dynamically typed, there are no constraints on the type of the elements of the lists.
This, on the one hand, gives more freedom to the user of the FreeCHR instance since it is easier to define rules on different types of constraints.
This is also possible when using Haskell, but needs additional work like defining and matching on a sum type.
On the other hand, in Python, one should always explicitly check if the constraints have the correct type in the head of the rule.

For example, the program in \autoref{fig::impl:gcd:py} allows values of any type in its state and will apply the rules only if integers are encountered.
Not checking the type of constraints beforehand is a serious source of runtime errors.
If a rule of the \python{gcd} program is applied to something other than two integers, the comparison operations (l. 3 and 6) in the guard or the subtraction (l. 7) in the body will raise an exception.
Such errors are impossible in the Haskell version in \autoref{fig::impl:gcd:hs} since the input type needs to be some sort of integer (\haskell{Integral}; see l. 1) which guarantees support for comparison (\haskell{Ord}) and subtraction (\haskell{Num}) \footnote{Both constraints are inherited via Haskell's type class \haskell{Real}.}.

\begin{figure}[t]
  \snippetHSsolverStep[firstnumber=1]
  \caption{Wrapped solver function type}
  \label{fig::impl:solver:hs}
\end{figure}

\begin{figure}[t]
  \subfigure[][Haskell]{%
    \begin{minipage}[b]{.9\textwidth}
      \snippetHSgcd[firstnumber=1]
    \end{minipage}
    \label{fig::impl:gcd:hs}
  }
  \subfigure[][Python]{%
    \begin{minipage}[b]{.9\textwidth}
      \snippetPYgcd[firstnumber=1]
    \end{minipage}
    \label{fig::impl:gcd:py}
  }
  \caption{Greatest common divisor}
  \label{fig::impl:gcd}
\end{figure}

\subsubsection{\textsc{rule}}
The implementations of \textsc{rule} can be seen in \autoref{fig::impl:rule}.
First, either function lazily computes the sequence of possible matchings (l. 5 in \autoref{fig::impl:rule:hs} and l. 3-5 in \autoref{fig::impl:rule:py}).
This is done using the \texttt{match} functions which are shown in \autoref{fig::impl:match}.
Both generate a lazy sequence of all permutations of the state that have the same length as the head of the rule (l. 3-4 and l. 3, respectively)
and select those for which the conjunction of the elementwise application (done by \haskell{zipWith} and \python{zip}) of the elements of the head evaluates to true,
\IE those for which $k_1(c_1) \band ... \band k_n(c_n) \band r_1(c_{n+1}) \band ... \band r_m(c_{n+m}) \equiv_{\tbool} \btrue$,
for a head $\left[k_1, ..., k_{n}, r_1, ..., r_{m}\right]$ and a permutation $\left[c_1, ..., c_{n+m}\right]$ of the state.
Then, both implementations of \textsc{rule} take the first of those sequences for which the guard $g$ holds, \IE $g(c_1,...,c_{n+m}) \equiv_{\tbool} \btrue$.
In total, this corresponds to l. \ref{alg::freechr:imp:set:alg:very-abstract:1} in \autoref{alg::freechr:imp:set:alg:very-abstract}.
If such a matching is found, the new state is computed and returned as in l. \ref{alg::freechr:imp:set:alg:very-abstract:2} (l. 7 and 8-11).
In the Python implementation, an explicit copy of the state is created to not modify the original state.
If no matching is found, the state is returned unchanged (l. 8 and 6-7).

Note that both variants explicitly create and return a function after getting the rule elements passed.
The implementations of \textsc{compose} (\autoref{fig::impl:compose}) work in the same way.

Also, the Python implementation makes heavy use of the \emph{spread operator} (\python{*} before function arguments) (l. 4 and 11).
In consequence, we can write the guard and body of a rule as functions that accept as many arguments as the head has patterns.
In Haskell, guard and body accept a list as their single argument (compare lines 6 and 7 in either implementation in \autoref{fig::impl:gcd}).

An issue specific to Python can be seen in \autoref{fig::impl:alldiff:py}.
Python does only allow expressions in \python{lambda} functions.
If any function of a rule needs to use statements, it must be defined separately.

\begin{figure}[t]
  \subfigure[][Haskell]{%
    \begin{minipage}[b]{.9\textwidth}
      \snippetHSrule[firstnumber=1]
    \end{minipage}
    \label{fig::impl:rule:hs}
  }
  \subfigure[][Python]{%
    \begin{minipage}[b]{.9\textwidth}
      \snippetPYrule[firstnumber=1]
    \end{minipage}
    \label{fig::impl:rule:py}
  }
  \caption{Implementations of \textsc{rule}}
  \label{fig::impl:rule}
\end{figure}

\begin{figure}[t]
  \subfigure[][Haskell]{%
    \begin{minipage}[b]{.9\textwidth}
      \snippetHSmatch[firstnumber=1]
    \end{minipage}
    \label{fig::impl:match:hs}
  }
  \subfigure[][Python]{%
    \begin{minipage}[b]{.9\textwidth}
      \snippetPYmatch[firstnumber=1]
    \end{minipage}
    \label{fig::impl:match:py}
  }
  \caption{Lazy matching}
  \label{fig::impl:match}
\end{figure}

\subsubsection{\textsc{compose}}
\autoref{fig::impl:compose} shows the implementations of \textsc{compose}.
\autoref{fig::impl:compose:hs} implements the algorithm as the binary operator \haskell{(<>)} of the \haskell{Semigroup} type class.
First, the left subprogram will be applied to the state (l. 5).
If this has no effect, the second subprogram is applied (l. 7).
Otherwise, the result of the first application which must have altered the state, is returned (l. 8).
One can easily see that the implementation is associative, as required by \haskell{Semigroup}.
In nested expressions like \haskell{p1 <> p2 <> ... <> pn}, the algorithm will always perform a depth-first search, beginning from the leftmost subprogram, effectively traversing the program from left to right.

The same is true for the Python variant in \autoref{fig::impl:compose:py}.
Here, we loop through the subprograms, passed to the function \python{compose} (l. 4-7) until a program alters the state (l. 6 \& 7).
Finally, \python{result} is returned which is either the input state (see l. 3) if the program made it through the loop without finding an applicable subprogram, or the first result of an effective application (see l. 5).

Contrary to the Haskell variant, \textsc{compose} is implemented as a variadic function by use of the spread operator in Python.
The consequence for the user can be seen in \autoref{fig::impl:gcd}, where we use \haskell{(<>)} to compose the rules (l. 4) in Haskell and \python{compose} in Python.

\begin{figure}[t]
  \subfigure[][Haskell]{%
    \begin{minipage}[b]{.9\textwidth}
      \snippetHScompose[firstnumber=1]
    \end{minipage}
    \label{fig::impl:compose:hs}
  }
  \subfigure[][Python]{%
    \begin{minipage}[b]{.9\textwidth}
      \snippetPYcompose[firstnumber=1]
    \end{minipage}
    \label{fig::impl:compose:py}
  }
  \caption{Implementations of \textsc{compose}}
  \label{fig::impl:compose}
\end{figure}

\subsubsection{\textsc{run}}
\autoref{fig::impl:run} shows the implementations of \textsc{run}.
The implementation in \autoref{fig::impl:run:hs} first applies the passed program to the query (\AKA the initial state) (l. 6).
If this has no effect, the state is returned (l. 3). 
Otherwise, \haskell{run} is recursively called with the new state (l. 4), repeatedly applying the program until exhaustion.

The algorithm in \autoref{fig::impl:run:py} accomplishes the same by the loop in lines 5 to 9,
which is interrupted if the state remains unchanged after applying the program (l. 6 \& 7).

\begin{figure}[t]
  \subfigure[][Haskell]{%
    \begin{minipage}[b]{.9\textwidth}
      \snippetHSrun[firstnumber=1]
    \end{minipage}
    \label{fig::impl:run:hs}
  }
  \subfigure[][Python]{%
    \begin{minipage}[b]{.9\textwidth}
      \snippetPYrun[firstnumber=1]
    \end{minipage}
    \label{fig::impl:run:py}
  }
  \caption{Implementations of \textsc{run}}
  \label{fig::impl:run}
\end{figure}

Again, the use of the spread operator in Python causes a minor difference in usage.
While in Haskell, we need to pass a list to \haskell{run}, in Python \python{run} it is a variadic function.
Hence, to calculate the greatest common divisor of $6$, $9$ and $12$, we call \haskell{run gcd [6, 9, 12]} in Haskell and \python{run(gcd)(6, 9, 12)} in Python and get in both cases \texttt{[3]} as the result.

\subsection{Side effects}
A major difference between the two presented implementations is the handling of side effects.
Python allows for arbitrary side effects like I/O, assigning variables or raising exceptions, as can be seen in \autoref{fig::impl:alldiff:py}.
Although this is also possible in Haskell (see \autoref{fig::impl:alldiff:hs}), it is highly discouraged by the language.
The programs raise an error if they encounter the same value in the query twice.
Although the implementations allow it, this behavior is not defined by the operational semantics and domain of future work.

\begin{figure}[t]
  \subfigure[][Haskell]{%
    \begin{minipage}[b]{.9\textwidth}
      \snippetHSalldiff[firstnumber=1]
    \end{minipage}
    \label{fig::impl:alldiff:hs}
  }
  \subfigure[][Python]{%
    \begin{minipage}[b]{.9\textwidth}
      \snippetPYalldiff[firstnumber=1]
    \end{minipage}
    \label{fig::impl:alldiff:py}
  }
  \caption{Implementation of \texttt{all\_diff} constraint}
  \label{fig::impl:alldiff}
\end{figure}

  \section{Limitations}\label{sec::freechr:limitations}
We want to briefly discuss the current limitations of FreeCHR as it is presented in this paper.

The very abstract operational semantics $\oma$ for CHR is meant as a tool for formal reasoning and an abstract baseline.
It does not describe the behavior of an actual implementation, since it is heavily non-deterministic.
As discussed above, it does also not account for trivial non-termination due to repeated application of a propagation rule on the same constraints.
Since the very abstract operational semantics for FreeCHR $\omstara$ is a direct translation of $\oma$, it inherits these issues.

The presented implementations are hence only defined on programs without propagation rules as they would inevitably get stuck on the repeated application of such a rule.
This issue is solved by the \emph{refined} operational semantics $\omega_r$ as presented by \cite{duck2004refined}.
Translating $\omega_r$ to FreeCHR is currently ongoing work.
For now, the implementations hence only serve as a proof of concept of the applicability of our framework and are not meant for actual use\footnote{Aside maybe from very small applications.}. 
We therefore did also not concern ourselves with applying any optimization techniques and do hence not yet expect any reasonable performance.

The approach presented herein is strictly to be understood as the formal foundation upon which we will build our future work, as it connects our framework to the original definition of CHR on the most fundamental level.

  \section{Related work}\label{sec::freechr:related-work}
\subsection{Constraint Handling Rules}
\subsubsection{Embeddings}
The first approach to embed CHR into a host language was via source-to-source transformation.
\cite{holzbaur1998compiling,holzbaur2000prolog} and \cite{schrijvers2004leuven}, \EG translate CHR via Prolog's macro system.
Similarily, \cite{abdennadher2002jack} and \cite{vanweert2005leuven} use a precompiler to translate CHR programs into Java,
\cite{wuille2007cchr} into C, \cite{nogatz2018chr} into Javascript, and \cite{barichard2024chr} into C++.

\cite{vanweert2008chr} introduces compilation schemes for imperative languages, upon which, \EG \cite{nogatz2018chr} builds.
The work of \cite{vanweert2010efficient} also provides compilation schemes for imperative languages, but optimizes the matching process by performing it lazily.
The algorithms do not first compute a full matching and check it against the patterns and the guard but collect it sequentially.

The inherent disadvantage of an approach using source-to-source compilation is the need for a precompiler in the build chain if the host language does not have a sufficiently expressive macro system like Prolog or LISP.
It comes with the cost of more sources of errors and an additional dependency that is often rather tedious to fulfill.
\cite{hanus2015chr} approaches this problem by extending the Curry compiler to be able to compile CHR code.
The obvious problem with this approach is that it is in most cases not viable to extend the compiler of the host language.

A relatively new approach by \cite{ivanovic2013implementing} was to implement CHR as an internal language in Java.
This has a major advantage: CHR can now simply be imported as a library, similarly as if implemented by a macro system like by \cite{schrijvers2004leuven}.
\cite{wibiral2022javachr} further builds upon this idea and introduces the idea of explicitly composing CHR programs of singular rules by an abstract and modular execution strategy and describing rules through anonymous functions.
This was our main inspiration and FreeCHR aims to improve and generalize this idea for arbitrary languages.

\subsubsection{Operational semantics}
\cite{duck2004refined} first formalized the behavior of existing implementations which were mostly derived from but not exactly true to existing formal definitions of operational semantics.
The techniques used in the implementations were also generalized and improved upon by \cite{vanweert2008chr} and \cite{vanweert2010efficient}.
\cite{duck2005compilation} standardized call-based semantics for CHR, especially in logical programming languages.

\subsection{Algebraic language embeddings}
The idea of algebraic embeddings of a domain specific language (DSL) into a host language was first introduced by \cite{hudak1998modular}.
The style in which the languages are embedded was later called \emph{tagless}, as it does not use an algebraic data type, to construct an abstract syntax tree (see, \EG \cite{carette2007finally}).
The \emph{tags} are the constructors of the data type which defines the syntax of the language.
Instead, the embedding directly defines the functions which implement the semantics of the language.
This is, defining the free $F$-algebra versus defining the concrete $F$-algebras, for a functor $F$ which defines the syntax of the language.
The advantage of embedding a DSL in this way is that it does not rely on external build tools, but can instead be easily implemented and used as a library.
It also enables the use of any features the host language offers, without any additional work.
With an external code-to-code compiler, it would be necessary to re-implement at least the syntax of any desired host language features.

\cite{hofer2008polymorphic} and \cite{hofer2010modular} then extended this idea by using type families in order to provide more flexibility concerning the semantics of the language.

\subsection{Logic and constraint based languages and formalisms}
CHR was initially designed as a tool to implement constraint solvers for user-defined constraints.
Hence, its domain intersects with those of \emph{Answer Set Programming} (ASP) and \emph{Constraint Logic Programming} (CLP).
However, on the one hand, CHR can rather be understood as a tool in combination with ASP or especially CLP, as it provides an efficient language to implement solvers, than as an alternative approach.
On the other hand, CHR has already exceeded its original purpose and developed more towards a general purpose language.

Another relevant logic based language is \emph{miniKanren} (see \cite{byrd2009relational}).
\emph{miniKanren} is a family of EDSLs for relational and logic programming.
There exists a myriad of implementations for plenty of different host languagesas well as formal descriptions of operational semantics (see \EG \cite{rozplokhas2020certified}).
The \emph{miniKanren} language family, though, seems to suffer from similar issues as traditional \emph{Constraint Handling Rules} does.
As there is no unified embedding scheme, there is an inherent disconnect between any implementation and the formally defined semantics.
Applying an approach similar to ours might be beneficial to the \emph{miniKanren} project as well, especially as it generally is implemented in a way similar to the methods described by \cite{hudak1998modular} or \cite{carette2007finally}.

  \section{Future Work}\label{sec::freechr:future-work}
With this foundational introduction of FreeCHR in place, our future work will focus on three main aspects.

First, since the execution algorithm presented here is only of limited practical use, we plan to introduce an algorithm that implements the \emph{refined} operational semantics $\omega_r$ of CHR.
The $\omega_r$ semantics of CHR is the basis for most existing implementations. Since we already have the \emph{very abstract} operational semantics as a baseline, we can use it as a basic anchor to verify any execution algorithm.
The algorithm we intend to formalize is already implemented in Haskell\footnote{\url{https://github.com/SRechenberger/free-chr-hs}} and Python\footnote{\url{https://github.com/SRechenberger/freechr-python}}.

Second, we currently work on translating the \emph{refined} operational semantics and verifying them against $\omstara$, as well as the aforementioned execution algorithm.
This will provide a base for theory concerning practically relevant aspects of FreeCHR.

We plan to verify both the execution algorithm implementing $\omega_r^{\star}$ and $\omega_r^{\star}$ itself by showing that they are valid concretization of $\omstara$.

Finally, we plan to tackle the issue of side effects that we encountered in the case study in \autoref{sec::freechr:instance:very-abstract:case-studies}.
This can be done by generalizing the functor $\chr_C$ to use monads to encapsulate arbitrary side effects (see \cite{jones1993imperative}).
This was already done in implementations in Haskell but is not yet formalized.
This is an important aspect, as it further formalizes practical aspects of everyday programming.
  \section{Conclusion}\label{sec::freechr:conclusion}
In this paper, we introduced the framework FreeCHR that formalizes the embedding of CHR, using initial algebra semantics.
We provided the fundamental definition of our framework which models the syntax of CHR programs over a domain $C$ as a \textbf{Set}-endofunctor $\chr_C$.
We defined the \emph{very abstract} operational semantics $\omstara$, using the free $\chr_C$-algebra $\chrstar$ and proved soundness and completeness \WRT the original \emph{very abstract} operational semantics $\omega_a$ of CHR.
We also provided a first abstract instance $\ExecA$ of FreeCHR, proved its correctness \WRT to $\omstara$ and discussed deterministic implementations of the algorithms in Haskell and Python.

Hereby, we established FreeCHR as a valid representation of the original definition of CHR and a usable framework for the implementation of the embedded language in a wide range of host languages.

Moreover, are the presented execution algorithm and the implementations of it, although very simple, to our knowledge the first implementations of CHR for which there is an actual prove of correctness \WRT to a formal definition of their operational semantics.
  \section*{Acknowledgments}
We thank our colleagues Florian Sihler and Paul Bittner for proofreading and their constructive feedback.

  \section*{Competing interests}
  The authors declare none.

  \bibliographystyle{class/tlplike}
  \bibliography{local/bibliography}
\end{document}